\documentclass[11pt,twocolumn]{article}

\usepackage[top=1.2in, bottom=1.2in, left=0.8in, right=0.8in]{geometry}
\usepackage{balance} 
\usepackage{graphicx} 
\usepackage{times}  
\usepackage{url} 
\usepackage{amsmath}

\usepackage{array}
\newcolumntype{L}[1]{>{\raggedright\arraybackslash}p{#1}}
\newcolumntype{R}[1]{>{\raggedleft\arraybackslash}p{#1}}

\usepackage{multirow}
\makeatletter
\def\url@leostyle{
  \@ifundefined{selectfont}{\def\UrlFont{\sf}}{\def\UrlFont{\small\bf\ttfamily}}}
\makeatother
\def\pprw{8.5in}
\def\pprh{11in}

\usepackage[pdftex]{hyperref}

\begin{document}

\title{Converging Work-Talk Patterns in Online Task-Oriented Communities}

\author{
Qi Xuan$^{1,2}$, Premkumar T Devanbu$^1$, Vladimir Filkov$^1$\\
\texttt{\{qxuan@, devanbu@cs., filkov@cs.\}ucdavis.edu}\\
$^1$Department of Computer Science, University of California, Davis, CA 95616, USA\\
$^2$Department of Automation, Zhejiang University of Technology, Hangzhou 310023, China
}

\date{}

\maketitle

\begin{abstract}
Much of what we do is accomplished by working collaboratively with others, and a large portion of  our lives are spent working and talking; the patterns embodied in  the alternation of working and talking can provide much useful insight into task-oriented social behaviors. The available electronic traces of the different kinds of human activities in online communities are an empirical goldmine that can enable the holistic study and understanding of these social systems. Open Source Software projects are prototypical examples of collaborative, task-oriented communities, depending on volunteers for high-quality work. Here, we use sequence analysis methods to identify the work-talk patterns of software developers in these online communities. 

We find that software developers prefer to persist in same kinds of activities, i.e., a string of work activities followed by a string of talk activities and so forth, rather than switch them frequently; this tendency strengthens with time, suggesting that developers become more efficient, and can work longer with fewer interruptions. This process is accompanied by the formation of community culture: developers' patterns in the same communities get closer with time while different communities get relatively more different. The emergence of community culture is apparently driven by both ``talk'' and ``work''. Finally, we also find that workers with good balance between ``work'' and ``talk'' tend to produce just as much work as those that focus strongly on ``work''; however, the former appear to be more likely to continue to be active contributors in the communities.
\end{abstract}

\vspace{1.4cm}

\section{Introduction}
A great deal of adult life is spent working. We work to create materials that fulfill human needs, to develop  advanced technologies, to govern, heal, and teach each other, etc. Our work is often collaborative, and often involves repeated activities: i.e., we commute, work, collaborate with others, etc. Collaborations involve both \emph{talking} and \emph{working}. We get some work done, talk with our colleagues to socialize, learn, or further co-ordinate tasks, and then work some more. The recurrent practices constitute patterns of activities that can be used to characterize  individuals, cluster them, and then predict their future behaviors; this has potential applications in various areas including crime control~\cite{koerin1978violent}, traffic forecasting~\cite{kitamura2000micro}, and marketing~\cite{bodapati2008recommendation}. In this paper, we will focus on the two most basic activities, i.e., work and talk. Note that, talking or communication, as important markers of human relations, play key roles in the coordination between pairs of co-operating individuals. As a result, they are commonly used to infer the social networks as the discrete spaces to study the dynamics of many other activities~\cite{newman2002email,onnela2007structure}. Here, however, we treat work and talk activities equally, and use sequence analysis methods to reveal the work-talk (W-T) patterns of the individuals in online task-oriented communities.

Sequence analysis, which has long history of being useful in molecular biology~\cite{watson1953the}, has been, as of recently, also used in social science~\cite{abbott1995sequence,brzinsky2006sequence}, where researchers investigate life courses~\cite{aisenbrey2010new}, and career trajectories~\cite{abbott1990measuring}. Whereas DNA sequences are curled up in three-dimensional space, social events are arranged according to their time of occurrence. Due to our interest in social phenomena mostly local in time, the positions of social events in a sequence refer to relative, rather than absolute, time points. In bioinformatics, a number of global and local sequence alignment methods are used to compare the molecules' genetic similarity within and across different organisms, so as to elucidate their biological functions~\cite{rosenberg2009sequence,sharma2008bioinformatics}. Here we adopt a local alignment method to find and enumerate short patterns in work-talk (W-T) sequences of different software developers. We use these short W-T pattern counts as data points for modeling developer behavior using hidden Markov models (HMMs)~\cite{eddy1996hidden}. The goodness of fit of these models are established via their ability to predict the numbers of larger patterns in the sequences~\cite{sharma2008bioinformatics}.

In collaborative projects the interplay of work and talk activities play a central role; therefore, the fitted parameters of these W-T HMMs can be used to characterize not just the W-T patterns of different individuals, but also help describe the projects' work culture. By ``work culture'' here we  mean the tendency of a group of individuals to share similar W-T patterns. The simplest such distinguishing pattern is that they either tend to work continuously
(thus creating shared work products) or talk continuously (co-ordinating work with others, and strengthening relationships). More complex patterns are a combination of the two. This connotation of ``culture'' is consistent with Etzioni's notion~\cite{etzioni1968the}: ``the set of assumptions shared by members of a societal unit which sets a context for its view of the world and itself''. It is known that community culture plays an important role in innovation~\cite{hurley1995group}, the quality of work-products~\cite{astrachan1988family}, and can facilitate the decision-making~\cite{chen1997reinforcing}. Community size and its evolution are related to productivity and efficiency.
Recent work on collaboration indicate that team size~\cite{wuchty2007the} and the team assembling mechanisms~\cite{guimera2005team} have significant effects on team performance.

We used open-source software (OSS) communities~\cite{dempsey2002who,mockus2002two,weber2004the} to study W-T patterns for three main reasons. First, the work in OSS communities is easy to observe, and most of the talk activities are meaningfully related (because of community norms) to work activities; this simplifies the observation of functional W-T patterns. Second, the work and talk activities in OSS communities are always archived~\cite{pattison2008talk}, so they are readily collected for analysis. Finally, performance properties, such as productivity, in terms of number of lines of code written, can also be measured using the state of the produced software. Communication is vital in making collaboration effective~\cite{xuan2014building}. Lack of communication between software developers may introduce more coordination problems~\cite{herbsleb1999architectures}, and thus the distributed developers do need to maintain awareness of one another to make the OSS projects successful~\cite{gutwin2004group}. A recent empirical research indicates that communication and commits may accelerate each other in Apache OSS projects~\cite{xuan2012measuring}, and similarly in Stack Overflow and GitHub~\cite{vasilescu2013stackoverflow}. However, it is also argued that, since both communication and working activities may compete for the time resources of individuals, communication may have some negative effect on the efficiency of work activities~\cite{mano2010mail,zelikovich2011negative}. In this case, communication can be considered as a kind of interruptions, recovering from which developers may take some time to continue their work~\cite{latoza2006maintaining,van1998interrupts}.    

\begin{table*}[!t]
\centering
\begin{tabular}{|p{1.6cm}|L{4.7cm}|p{3.7cm}|R{1cm}|R{1cm}|R{1.4cm}|R{1.2cm}|}
\hline
Project & Description & Time frame &\#Users & \#Devs & \#S-Devs & \#Files \\
\hline
Activemq & Integration patterns server & 2005/12/12-2012/03/16 & 2012 & 28 & 6 & 16788\\
\hline
Ant & Build tool & 2000/01/13-2012/03/16 & 1402 & 44 & 9 & 11620\\
\hline
Axis2\_c & Web services engine & 2004/02/03-2012/03/15 & 582 & 24 & 8 & 10262\\
\hline
Axis2\_java & Web services engine & 2001/01/30-2012/03/19  & 3738 & 72 & 15 & 129978\\
\hline
Camel & Integration framework & 2007/03/19-2012/03/17 & 805 & 31 & 6 & 36965\\
\hline
Cxf & Web services framework & 2005/07/22-2012/03/16 & 427 & 45 & 7 & 37867\\
\hline
Derby & Database management system & 2004/08/10-2012/03/22 & 1118 & 35 & 16 & 6563\\
\hline
Lucene &  Search software & 2001/09/11-2012/03/23 & 2102 & 41 & 14 & 6674\\
\hline
Mahout &  Machine learning library & 2008/01/15-2012/03/23 & 533 & 15 & 6 & 5123\\
\hline
Nutch &  Web search software & 2005/01/25-2012/03/22 & 556 & 16 & 6 & 3072\\
\hline
Ode & Web services & 2006/02/18-2012/03/22 & 365 & 17 & 6 & 11006\\
\hline
Openejb &  Container system and server & 2002/01/18-2012/03/22 & 169 & 38 & 5 & 43960\\
\hline
Solr &  Enterprise search platform & 2006/01/20-2011/03/01 & 825 & 19 & 8 & 8534\\
\hline
Wicket &  Web application framework & 2004/09/21-2012/03/21 & 539 & 24 & 8 & 48045\\
\hline
\end{tabular}
\caption{Basic properties of the fourteen OSS communities.}
\label{Tab:Properties}
\end{table*}

In this paper, by studying W-T patterns of developers in OSS communities, we make the following main contributions.
\begin{itemize}
\item We establish HMM models for the W-T sequences of developers and find the evidence of community culture: developers in the same community tend to have more similar W-T patterns than those from different communities, and this pattern-affinity strengthens with time. 

\item We observed that developers who have balanced W-T patterns are just as productive as those who work more continuously (fewer interruptions), but the former tend to stay active longer; this suggests that W-T balance is important to sustain OSS communities.

\item We create social and cooperation networks, and find that the convergence of W-T patterns between a given pair of developers appears to be re-inforced by both talk- and work-related interactions. This indicates that the emergence of a community W-T culture is apparently driven by both ``work'' and ``talk'', and thus offers a novel perspective on the co-evolving mechanisms of socio-technical, interdependent networks~\cite{brummitt2012suppressing,buldyrev2010catastrophic,xuan2013reaction}. 
\end{itemize}

\section{\label{Description}Work-Talk Activities in OSS}
Human activities can be represented by time-series denoting the event occurrences. As a generalization, different kinds of activities can be represented by an asynchronous multiple time-series~\cite{lutkepohl2005new} since people seldom do different things at exactly the same time;  this is quite different from the multiple time-series in economics where different kinds of indexes are recorded at the same time for comparison. Since work and talk are the activities of concern here, we will use sequence analysis  to study these two kinds of activities in OSS communities, with primary focus on their \emph{relative time orders}, without concern for the \emph{precise time of occurrence}.

\subsection{Data Description}
We collected the individual work and talk activities from 31 OSS developer communities in the \emph{Apache Software Foundation} on March 24th, 2012. In each community, there are several volunteer developers who contribute by committing to files, i.e., adding or removing software code; these activities are recorded in a Git repository.
These are our ``work events'', or W. Members of the online communities communicate using the \emph{developer mailing lists} with the rest of the community through emails to share programming knowledge, and coordinate with others. We record the sent emails as ``talk events'', or T, of a developer (the received emails are included in the talk activities of others). Using this data, a W-T sequence concerning the work and talk activities can be recorded for each developer. Note that messages may be automatically posted to a mailing list in an OSS community to inform others when some work is done. In order to exclude such trivial talk activities, here we just consider those response emails~\cite{xuan2012measuring} which takes up about 73\% of all messages~\cite{bird2006mining}. Moreover, we also use a semi-automatic approach to solve the problem of multiple aliases\cite{bird2006mining}. 

We pre-process the W-T sequence data in several ways. To ensure a sufficient number of samples to reliably compare the W-T patterns between the pairs of developers of the same or from different communities,  we select a subset of the developers with  sequences including at least 500 work and talk activities, and a subset of communities with at least 5 such developers. We acknowledge a risk of left-censorship of both work \& talk activities, if any OSS communities did not archive their emails, or if they had used different version control systems before they moved to Git, some early data could be lost. Besides, it is known that many individuals need to first earn social capital in the OSS project  by communicating with others before they are accepted as developers~\cite{bird2007open, gharehyazie2013social}. As a result, we often observe long, pure work (or talk) subsequences before the first talk (or work) activity of a developer. In this study, we remove these trivial prefixes of pure work or talk activities, i.e., we only consider W-T sequences starting from the first work (or talk) activity if it occurred after the first talk (or work) activity.
 
After pre-processing the data, we are left with 14 communities totally containing 120 developers, and several of their basic properties are presented in Table~\ref{Tab:Properties}. Note that besides the developers, we also list the number of active users (including developers) in each project. These users might not directly commit to files, but they may contribute to the communities by other ways, such as reporting bugs etc. 

\subsection{Pattern Analysis}
\begin{figure}[!t]
\includegraphics[width=\columnwidth]{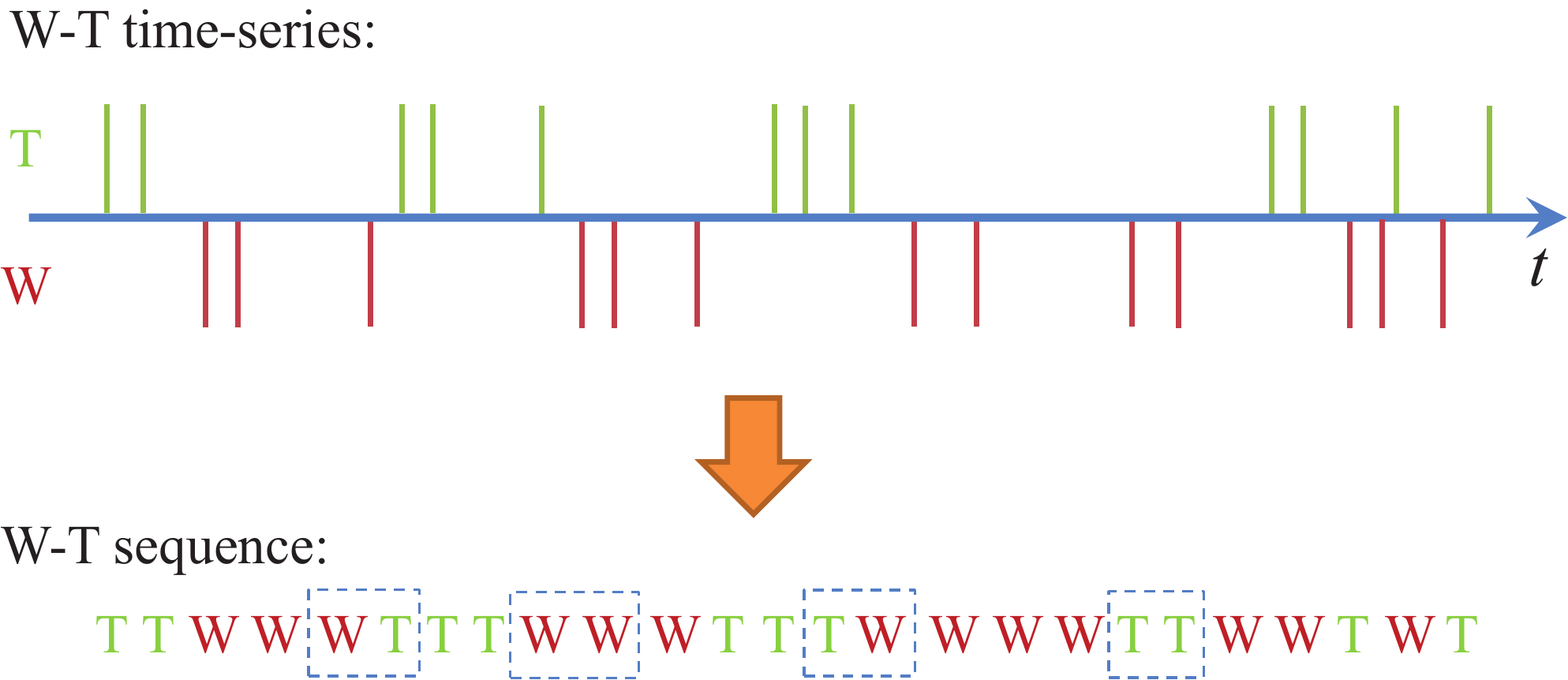}
\caption{A multiple time-series of work and talk activities and the corresponding W-T sequence. The four different two-patterns, i.e., WW, WT, TW, and TT, are marked by the dashed rectangles.}
\label{Fig:WTSequence}
\end{figure}

A $G$-pattern in a sequence over the alphabet $\{\textrm{W},\textrm{T}\}$ is a subsequence of length $G$. There are total $2^G$ possible different $G$-patterns. Typically, the length of the patterns is much shorter than the length of the given sequence. In our study we focus on 2-patterns and 3-patterns. Given a sequence $\theta=\{s_1,s_2,\ldots,s_h\}$ over $\{\textrm{W},\textrm{T}\}$, we count the occurrence of each of the $2^G$ patterns, by rolling a window of size $G$ over the sequence, and incrementing the count for the pattern we find. E.g., in the W-T sequence shown in Figure~\ref{Fig:WTSequence}, the four possible 2-patterns, WW, WT, TW, and TT,  occur eight, five, five, and six times, respectively.

\begin{figure}[!t]
\centering
\includegraphics[width=\columnwidth]{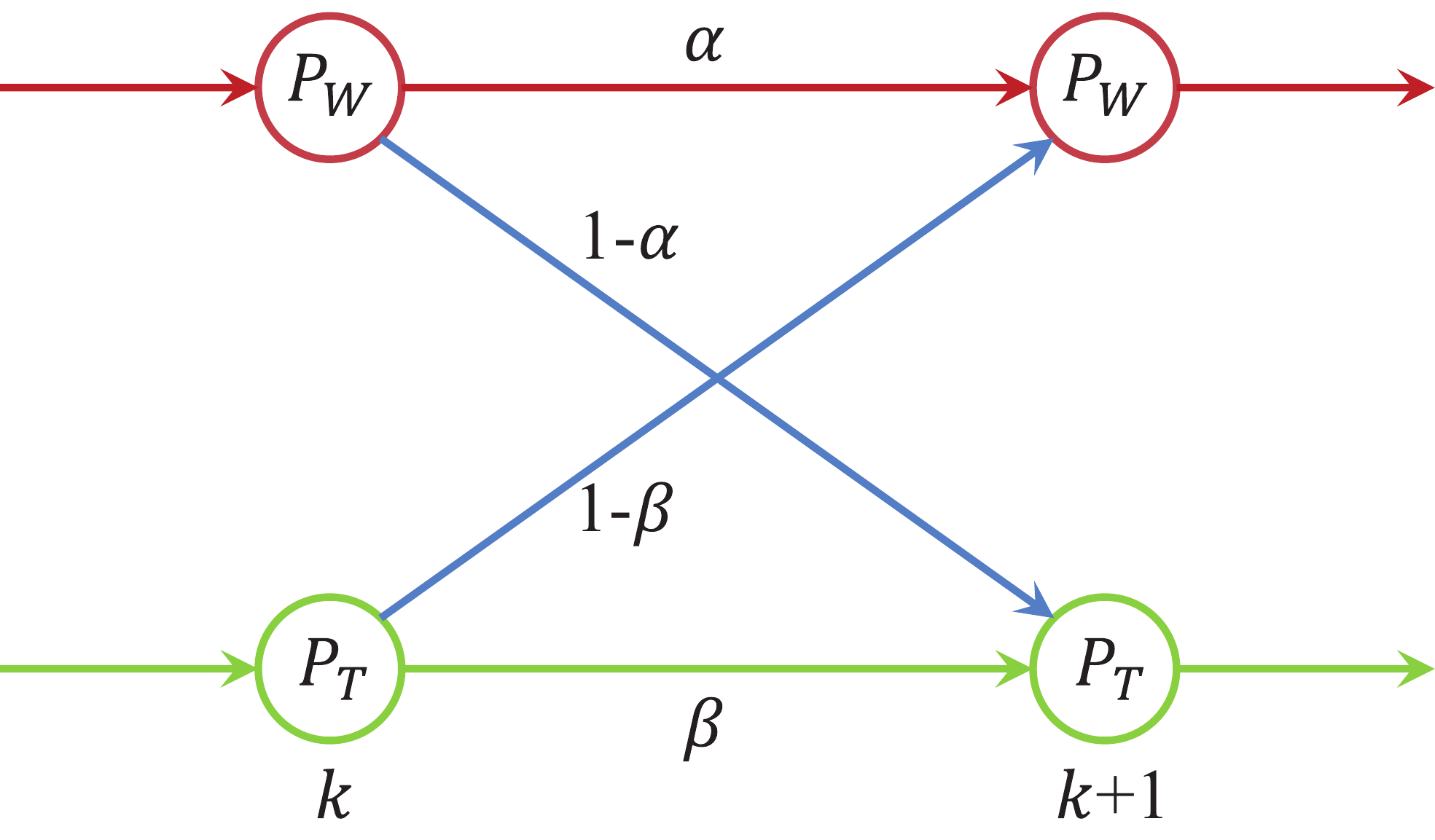}
\caption{An HMM with two states, i.e., ``work'' and ``talk'', denoted by ``W'' and ``T'', respectively. The model is used to explain the W-T patterns of developers in different communities.}
\label{Fig:MarkovModel}
\end{figure}

To assess the probability that a pattern occurs by chance, we create a null (baseline) model by randomizing the observed W-T sequence so as to preserve the proportion of work to talk activities. This can be achieved, e.g., by using the R package's~\cite{team2005r}, sample() function, on the sequence indexes. Then, the preference for pattern $P$ in the observed sequence, $\theta$, over the randomized sequence, $\theta^*$, is calculated by the relative difference between the counts for that pattern, $C_P$ and $C^*_P$, in the respective sequences,
\begin{equation}
\lambda_P=\frac{C_P-\langle{C^*_P}\rangle}{\langle{C^*_P}\rangle}\times{100\%}.
\label{Eq:Pattern:Significance}
\end{equation}
For each pattern $P$ in a sequence, we also calculate its Z-score~\cite{kashtan2005spontaneous} as $\lambda_P{\langle{C^*_P}\rangle}/\varsigma$, where $\varsigma$ is the standard deviation of the pattern counts in $\theta^*$. For $\langle{C^*_P}\rangle$, we generated 100 randomized sequences for each observed one, and the absolute values of the Z-scores for 2-patterns in our study are larger than 5, in 462 out of 480 cases, indicating that the observed counts are surprising. These random sequences are also used as references to evaluate the HMM model (represented in Figure~\ref{Fig:MarkovModel}) on predicting 3-patterns. (See the Appendix for further mathematical description of the HMM.)

\section{\label{Pattern}Work-Talk Patterns are Conserved over Short Time Intervals}
A work-talk pattern is a string, or word, comprised of the letters W and T, signifying the work and talk activities.
The simplest non-trivial pattern is of length two, which considers two successive activities (same or different) at a time. There are four possible different two-patterns: work-work (WW), work-talk (WT), talk-work (TW), and talk-talk (TT). Given an observed W-T sequence for each person, we count in it the occurrences of two-patterns. Then, we derive the preference of each pattern, denoted by $\lambda_i,i=1,2,3,4$, respectively, in the real sequences as compared to random ones (for each real sequence we bootstrapped from it 100 simulated W-T sequences, by randomizing the order of its elements). In our data set, we find that, on average for all developers, $\lambda_1=148.9\%$ and $\lambda_4=40.5\%$, while $\lambda_2=-38.0\%$ and $\lambda_3=-38.6\%$, i.e., WW and TT are positively enriched, while WT and TW are negatively enriched. This suggests that developers much prefer to persist with one activity-type, rather than switch between activities.

\begin{figure}[!t]
\centering
\includegraphics[width=\columnwidth]{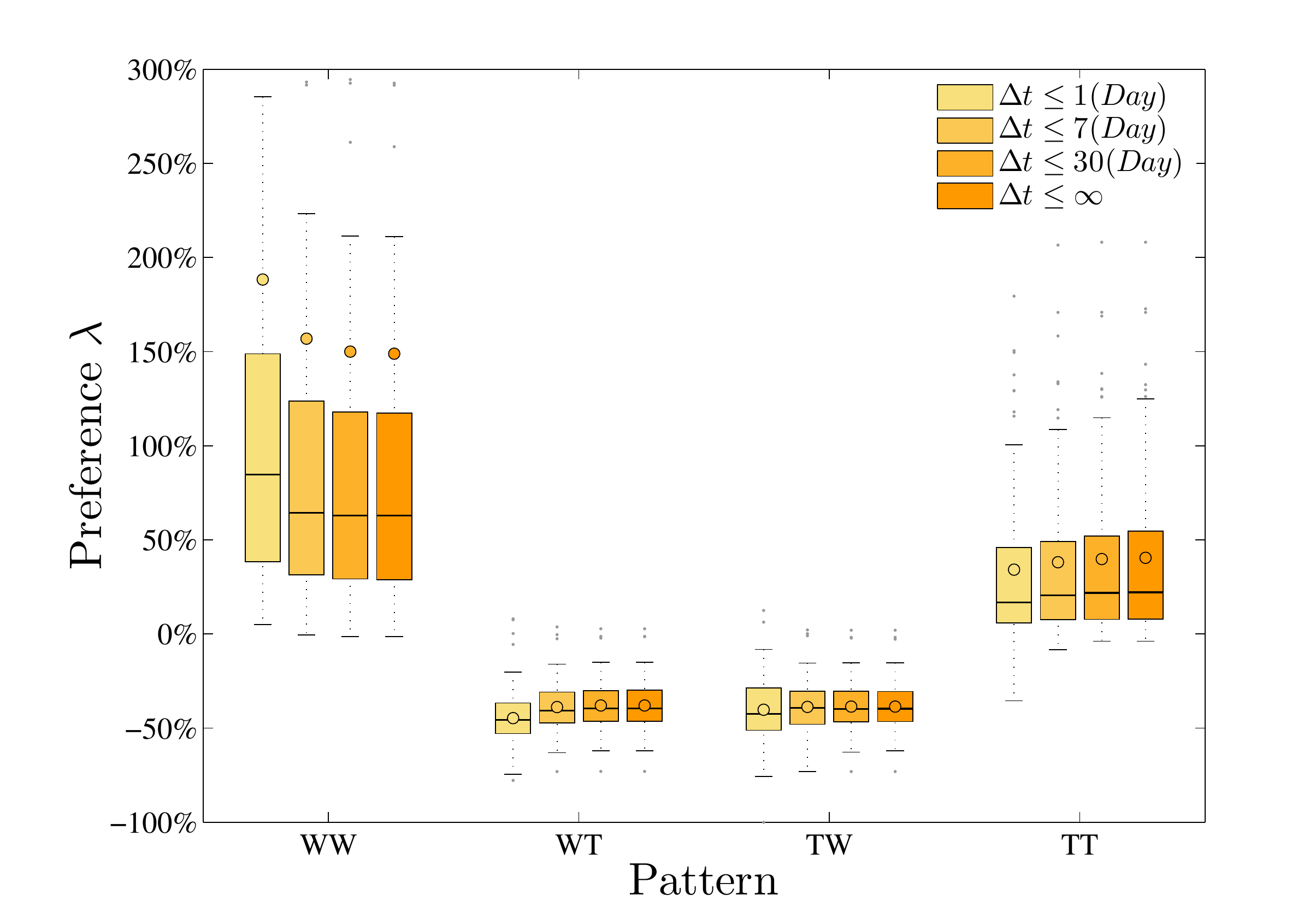}
\caption{The box-and-whisker diagram for the preferences of the four different two-patterns in the real W-T sequences under the different time-interval conditions by comparing with the random ones.}
\label{Fig:IEffect}
\end{figure}

It may be argued that two successive activities should not be considered as a two-pattern if the time interval between them is relatively long, e.g., longer than one month. To show that our method is robust with respect to time-scale, we also calculate the relative difference by varying the thresholds for the time-intervals over which we consider the two-patterns. We vary the thresholds, denoted by $\xi=1,7,30$ (days), and only the patterns with intervals $\leq{\xi}$ are considered. The results are shown in Figure~\ref{Fig:IEffect}, where we can see that WW and TT patterns are always much more preferred than WT and TW patterns in the real sequences under thresholds varying from one day to one month. Interestingly, we also find a slight trend that the WW pattern becomes more preferred, and the TT pattern less preferred, when we exclude more repeated activities with relatively shorter time intervals (and thus a smaller $\xi$). Since the number of these long time-interval patterns is relatively small (2.2\% and 0.3\% for $\xi=7$ and $\xi=30$, respectively), this slight trend still indicates that developers are more likely to start and end a repeated and relatively compressed work sequence with talk activities, viz., talk activities plays important role in enabling new tasks (work activities) in these online communities.

In order to study W-T patterns in more detail, we create a two-state HMM with parameters $\alpha$ and $\beta$ representing the conditional transition probabilities $P(\textrm{W}|\textrm{W})$ and $P(\textrm{T}|\textrm{T})$, respectively, for each developer (see the Appendix). Based on the model, we have
\begin{eqnarray}
\alpha=\frac{P_{\textrm{WW}}}{P_{\textrm{WW}}+P_{\textrm{WT}}}, \quad \beta=\frac{P_{\textrm{TT}}}{P_{\textrm{TT}}+P_{\textrm{TW}}},\label{Eq:Parameter}
\end{eqnarray}
where $P_{\textrm{WW}}$, $P_{\textrm{WT}}$, $P_{\textrm{TW}}$, and $P_{\textrm{TT}}$ denote the probabilities of the four different two-patterns for each developer, and can be estimated by the counting numbers of the four different two-patterns, i.e., $M_i,i=1,2,3,4$, respectively. Then, we get
\begin{eqnarray}
\alpha=\frac{M_1}{M_1+M_2}, \quad \beta=\frac{M_4}{M_3+M_4},\label{Eq:PEstimate}
\end{eqnarray}
as long as the corresponding W-T sequence contains enough elements. Here, it is not difficult to prove that the condition $|M_2-M_3|\leq{1}$ must be satisfied for any W-T sequence.

\begin{figure}[!t]
\centering
\includegraphics[width=\columnwidth]{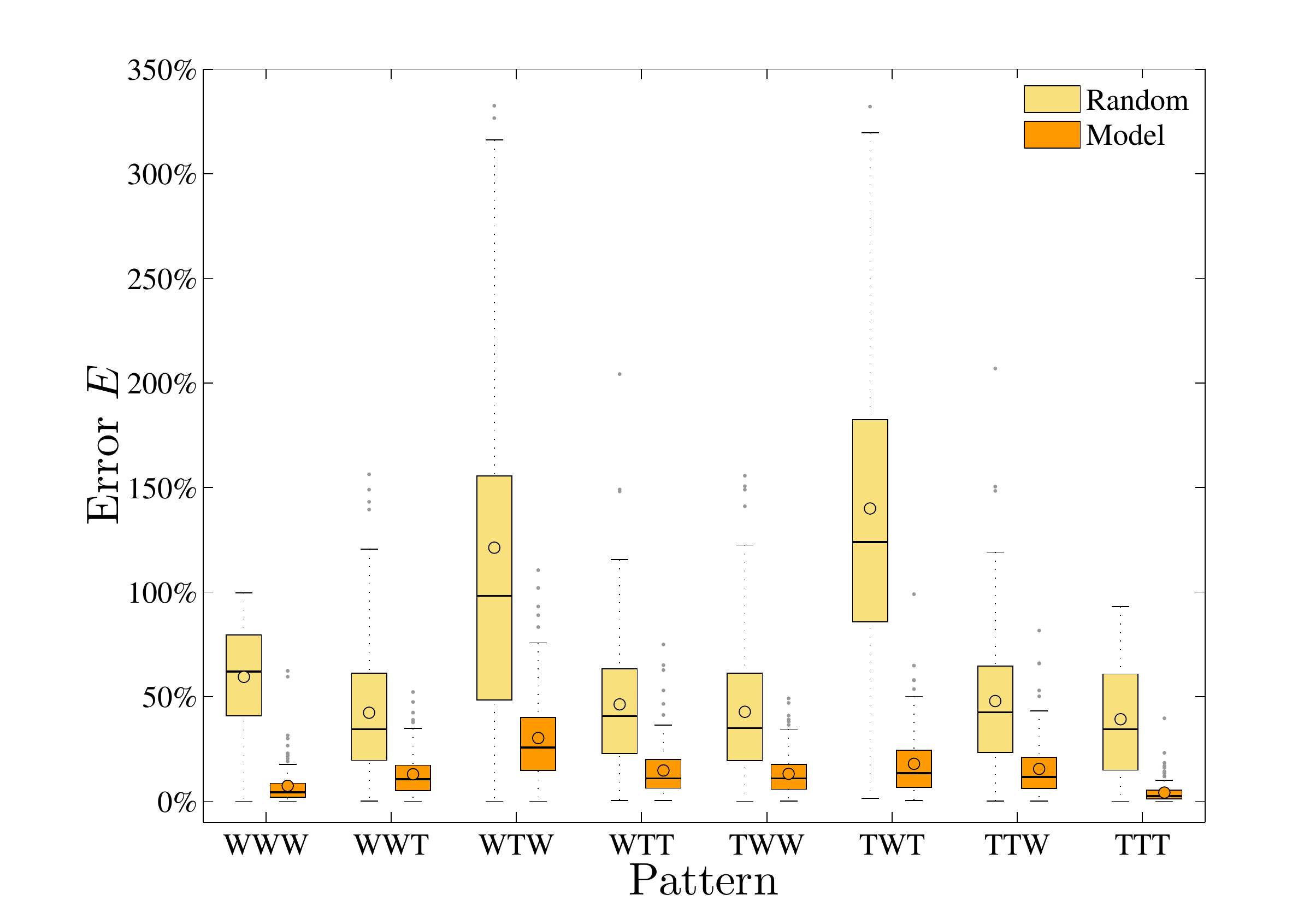}
\caption{The box-and-whisker diagram for the relative errors of the eight different three-patterns introduced by the random mechanism and the HMMs, comparing with the real ones.}
\label{Fig:ErrorBox}
\end{figure}

This HMM is fully determined by the numbers of the four different two-patterns. We validate the model by checking its ability to predict the numbers of larger patterns, e.g., three-patterns. There are eight different three-patterns, WWW $\ldots$ TTT. For each developer, we denote by $M_{i}$, $M^*_{i}$, and $M^\circ_{i}$, $i=1,2,\ldots,8$, the numbers of three-patterns in the real sequence, the random sequence, and the sequence created by the model (with the same length and the initial element), respectively. Then, for this developer, we can calculate the relative error introduced by the random mechanism and the model by
\begin{eqnarray}
E^*_{i}=\frac{|M^*_{i}-M_{i}|}{M_{i}}, \quad E^\circ_{i}=\frac{|M^\circ_{i}-M_{i}|}{M_{i}},\label{Eq:Predict:Error}
\end{eqnarray}
respectively. For each developer, and for each three-pattern, the difference between these two errors can be used to validate the ability of the HMM in predicting that pattern: viz., if the relative errors introduced by the model is significantly smaller than those introduced by the random mechanism, it is reasonable to believe that the model feasibly predicts that pattern. The differences between these two kinds of errors for all the eight different three-patterns are visualized by the box-and-whisker diagrams shown in Figure~\ref{Fig:ErrorBox} with all the developers considered together, where we can see that the HMM does indeed predict the numbers of all the eight three-patterns with significantly smaller relative errors ($p=1.8\times10^{-16}$ on average) than the random mechanism for the developers we studied, i.e., 14.5\% versus 67.4\% on average.

\section{\label{Community}Community Culture}
Generally, people sharing similar habits and interests are more likely to come together; once together, they could further influence each other, so as to form the community culture. Then, one interesting question is:  do  developers in the same OSS communities present more similar W-T patterns (or closer $\alpha$ and $\beta$ more specifically) than  developers from different communities, i.e., can W-T patterns be used as a metric to characterize community culture?

\subsection{\label{Culture}Converging Work-Talk Patterns and Community Culture}

\begin{figure}[!t]
\centering
\includegraphics[width=\columnwidth]{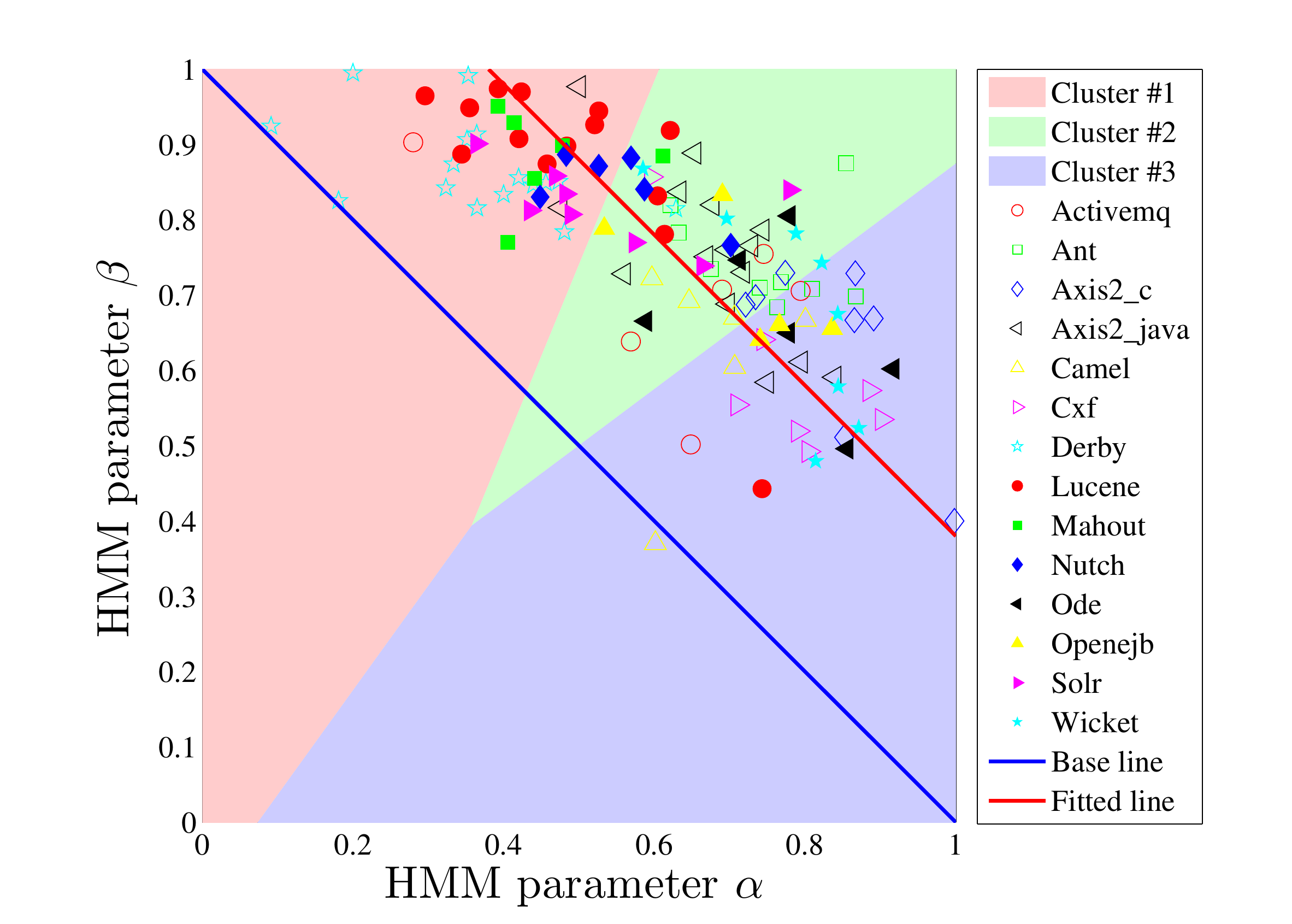}
\caption{Visualization of developers on $\alpha$-$\beta$ plane by considering their whole sequences, where developers are point and those of the same communities are marked by the same symbols. The parameters are grouped into three clusters by the ``K-means'' method. The base line is formed by the HMM parameters of the random W-T sequences with different fractions of work activities. The points are fitted by the linear function $\alpha+\beta=\varepsilon$, with $\varepsilon=1.38$.}
\label{Fig:PFigure}
\end{figure}

In order to answer the above question, we first visualize all $(\alpha,\beta)$ pairs in the $\alpha-\beta$ plane, as shown in Figure~\ref{Fig:PFigure}, where the developers of the same communities are marked by the same symbols.  Evidence of clustering is visually apparent: the points representing the developers in the same communities are indeed closer to each other when compared with those from different communities. We further classified all the developers into three groups by the k-means method~\cite{jain1999data}, and find that most  developers in the same communities are centralized in one of three clusters, rather than uniformly distributed in all the three, which indicates different community cultures that emphasize continuous work (cluster \#3), talk (cluster \#1), or both (cluster \#2), respectively. Here, we also provide the baseline formed by the HMM parameters of the W-T sequences that are generated by the random mechanism with different fractions of work activities. Since this baseline must satisfy $\alpha+\beta=1$, and almost all the points being above this based line validates again the preferred
patterns WW and TT in all communities.

\begin{figure}[!t]
\centering
\includegraphics[width=\columnwidth]{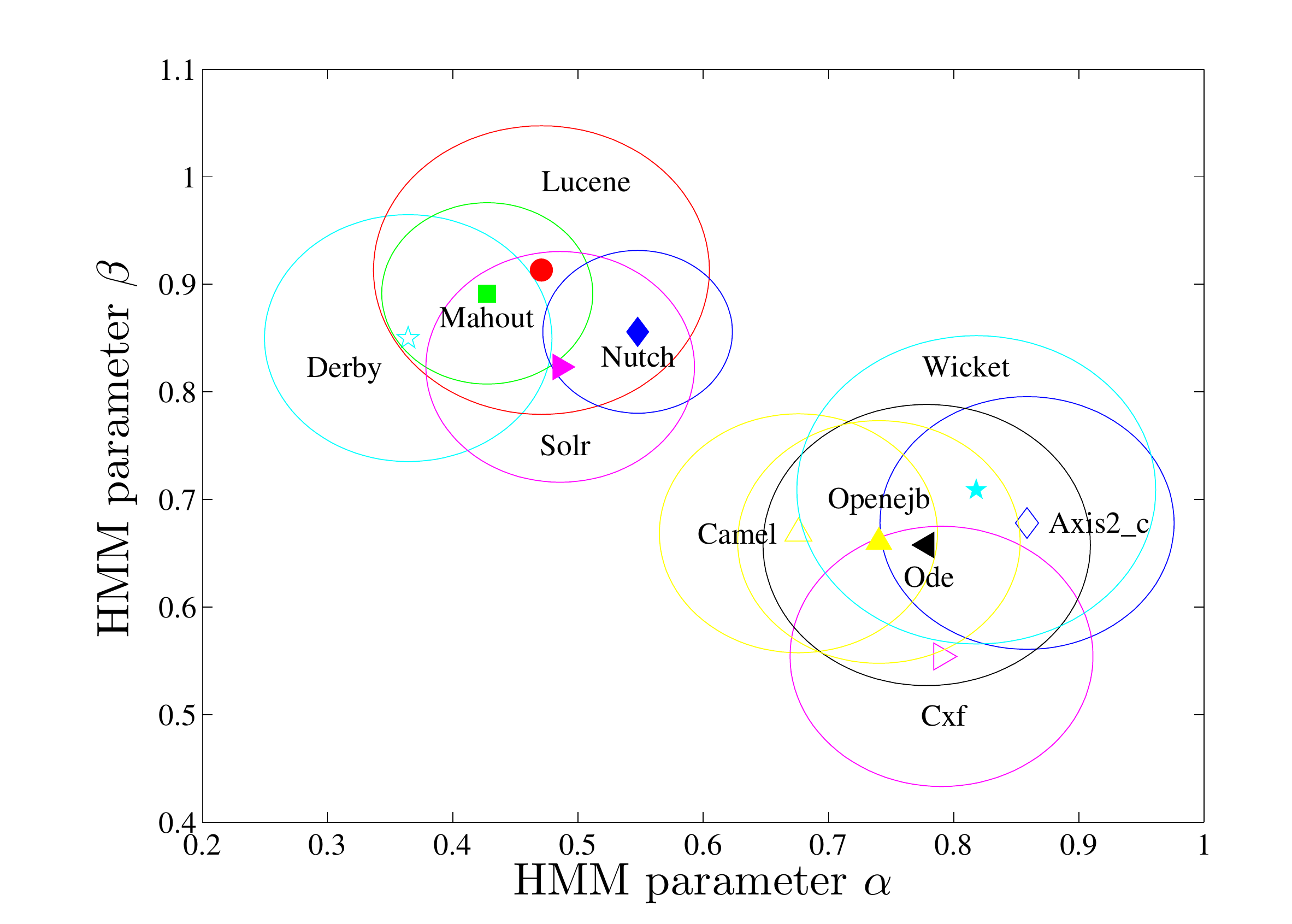}
\caption{The centers and the respective diversities (the large circles) of the eleven communities on $\alpha-\beta$ plane, defined as the medians of the HMM parameters of their developers and the average distances of HMM parameters between the developers and the corresponding centers, respectively.}
\label{Fig:PJDomain}
\end{figure}

More specifically, most developers ($\geq{50\%}$) in \emph{Derby}, \emph{Lucene}, \emph{Mahout}, \emph{Nutch}, and \emph{Solr} belong to cluster \#1, which corresponds to mostly talk activities (high $\beta$), while most of the developers in \emph{Axis2\_c}, \emph{Camel}, \emph{Cxf}, \emph{Ode}, \emph{Openejb}, and \emph{Wicket} belong to cluster \#3, corresponding to mostly work activities (high $\alpha$). As a whole, we define the center of a community in $\alpha-\beta$ plane by the median of the HMM parameters of the developers in it, then calculate its diversity by the average distances of HMM parameters between the developers and the center, as shown in Figure~\ref{Fig:PJDomain} for the above 11 communities. It is interesting to find that the communities sharing similar W-T patterns also belong to similar domains (description in Table \ref{Tab:Properties}). For example, \emph{Lucene}, \emph{Nutch}, and \emph{Solr} are all about ``search'' and they are intrinsically related to each other, just like the introduction of \emph{Nutch} on its web site: ``Stemming from Apache \emph{Lucene}, Apache \emph{Nutch} now builds on Apache \emph{Solr} adding web-specifics''. Besides, \emph{Axis2\_c}, \emph{Cxf}, and \emph{Ode} are all about ``services'', while each of \emph{Camel}, \emph{Cxf}, and \emph{Wicket} is a software framework that provides a shared architecture for class of applications.

More formally, if we denote by $\alpha_i$ and $\beta_i$ the HMM parameters of developer $d_i$, we can calculate the Euclidean distance of HMM parameters between two developers $d_i$ and $d_j$ by
\begin{equation}
\rho_{ij}=\sqrt{(\alpha_i-\alpha_j)^2+(\beta_i-\beta_j)^2},\label{Eq:EuclideanDistance}
\end{equation}
as a quantitative metric for the similarity between the W-T patterns of developers, i.e., the shorter the distance between them, the more similar the W-T patterns of the two developers. Then, we compare the distances of HMM parameters between all pairs of developers in the same communities with those between pairs of  developers from different communities, and find that the former list of distances are significantly shorter ($p=0$) than the later ones. These qualitative and quantitative analysis lend support to using the HMM parameters as a reasonable proxy for the way the interplay of work and talk testify to community culture.

\begin{figure}[!t]
\centering
\includegraphics[width=\columnwidth]{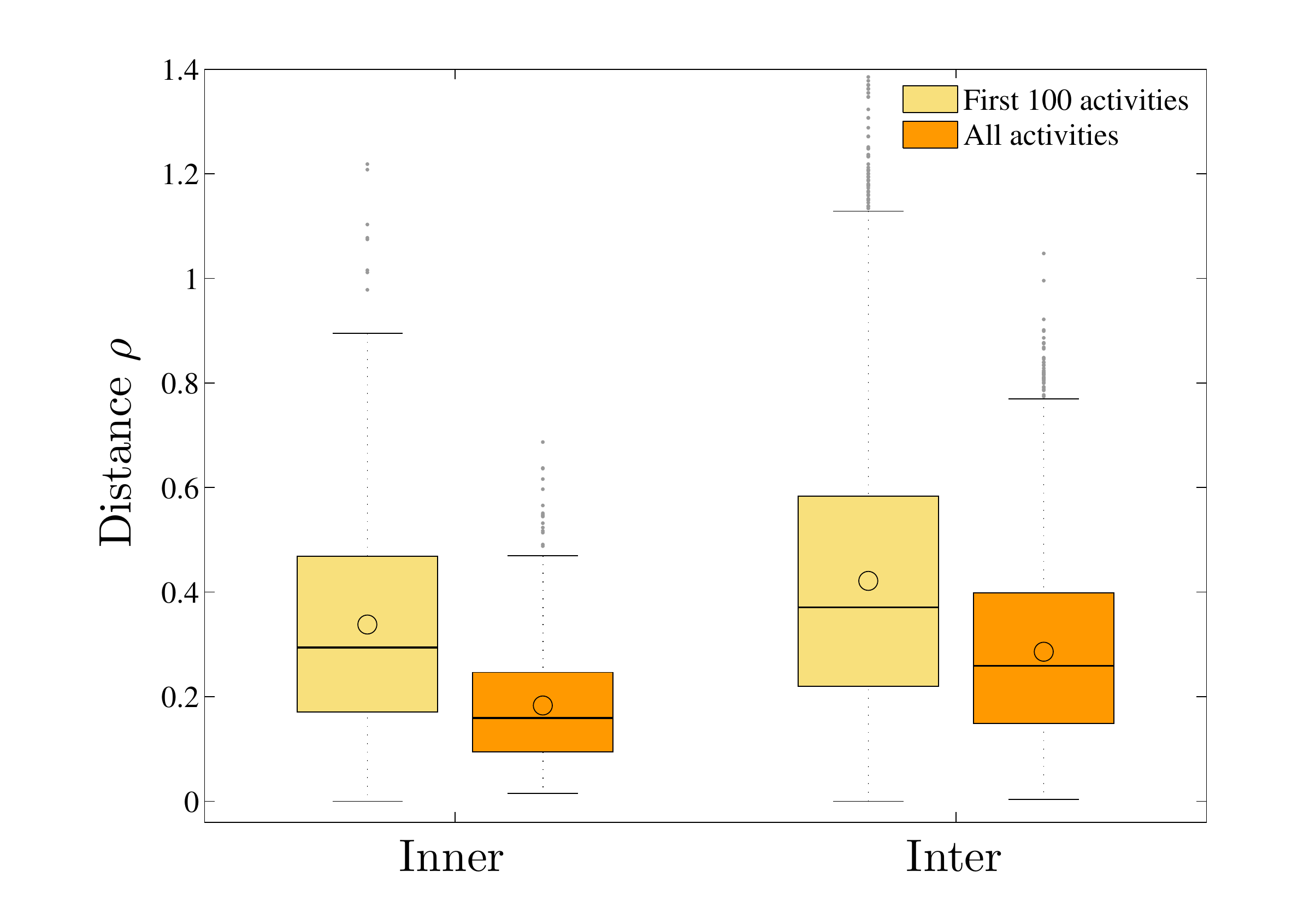}
\caption{The box-and-whisker diagrams for the distances of the HMM parameters of the first 100 activities and those of the whole W-T sequences between pairs of developers inner and inter communities.}
\label{Fig:DBox}
\end{figure}

\begin{figure}[!t]
\centering
\includegraphics[width=\columnwidth]{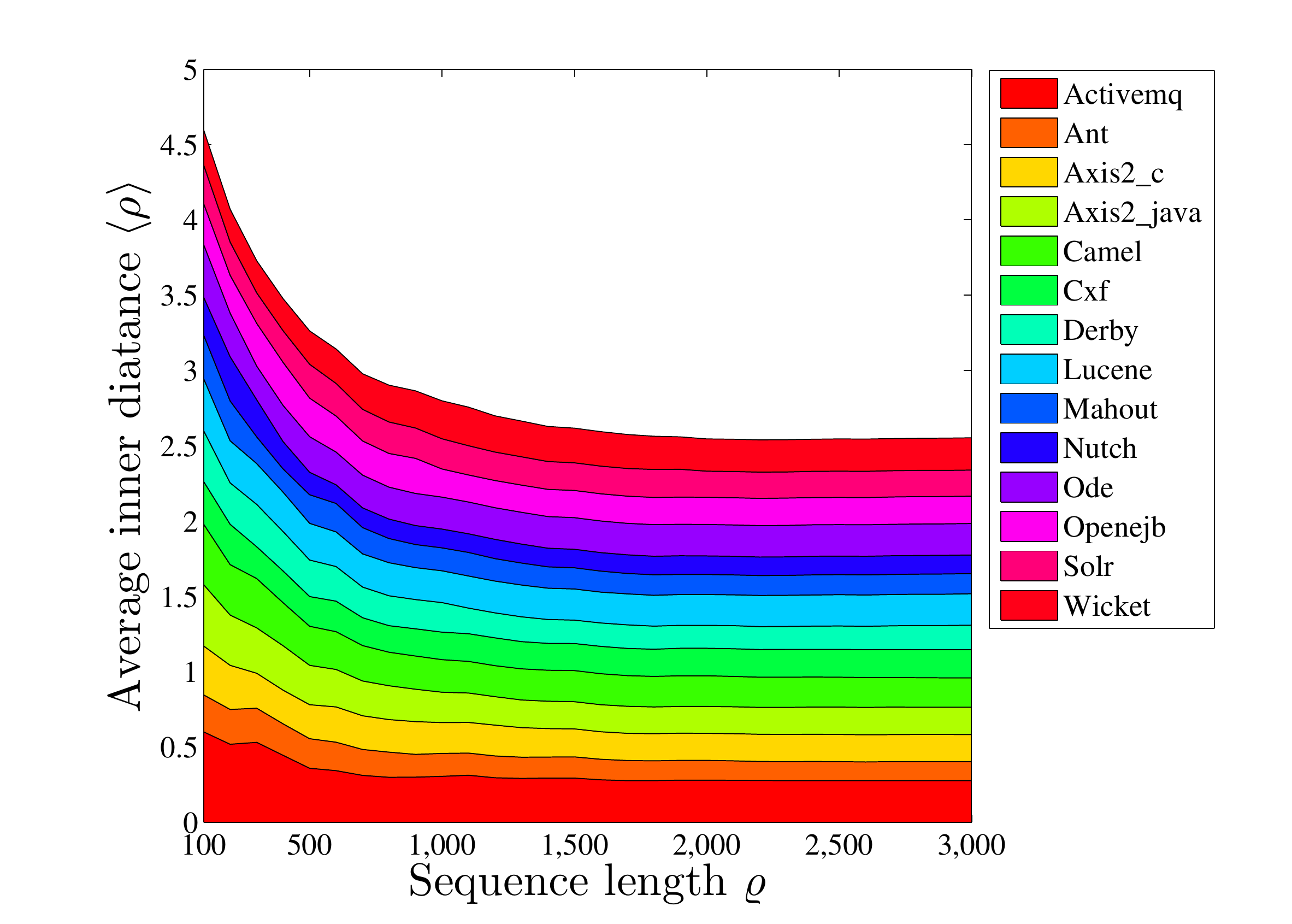}
\caption{The average inner distances of HMM parameters between pairwise developers for the fourteen communities.}
\label{Fig:PDCurve}
\end{figure}

\begin{figure}[!t]
\centering
\includegraphics[width=0.48\columnwidth]{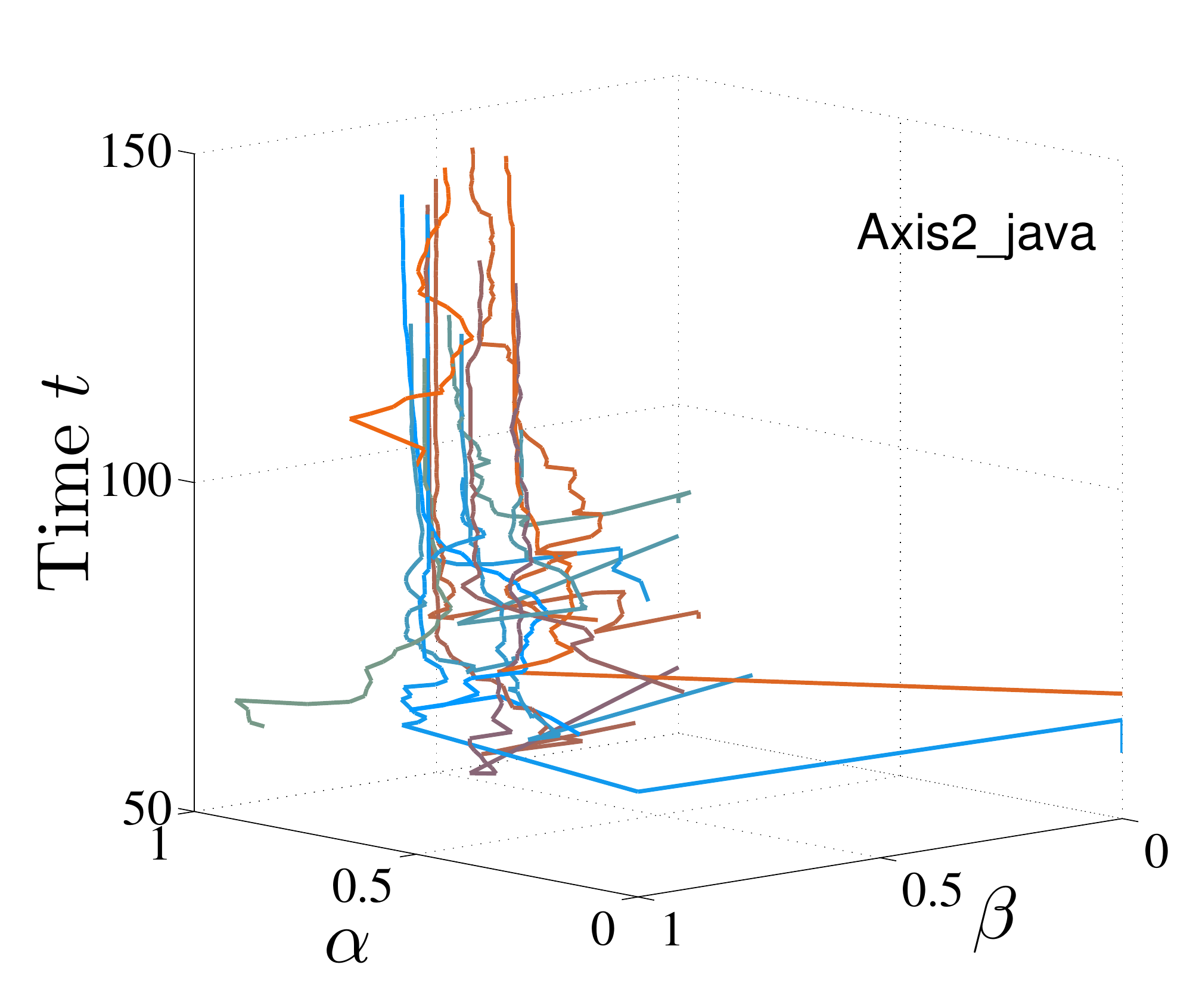}
\includegraphics[width=0.48\columnwidth]{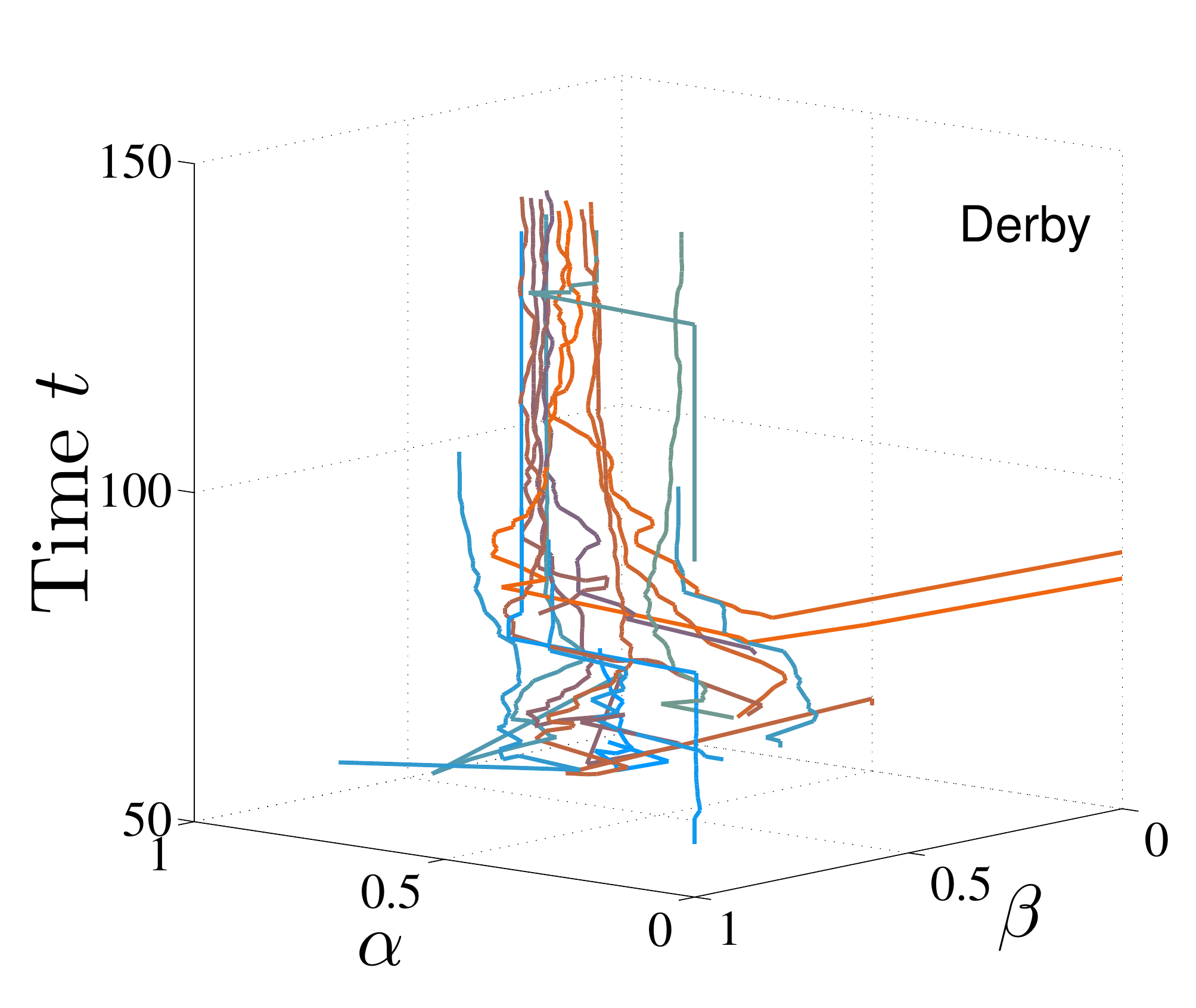}\\
\includegraphics[width=0.48\columnwidth]{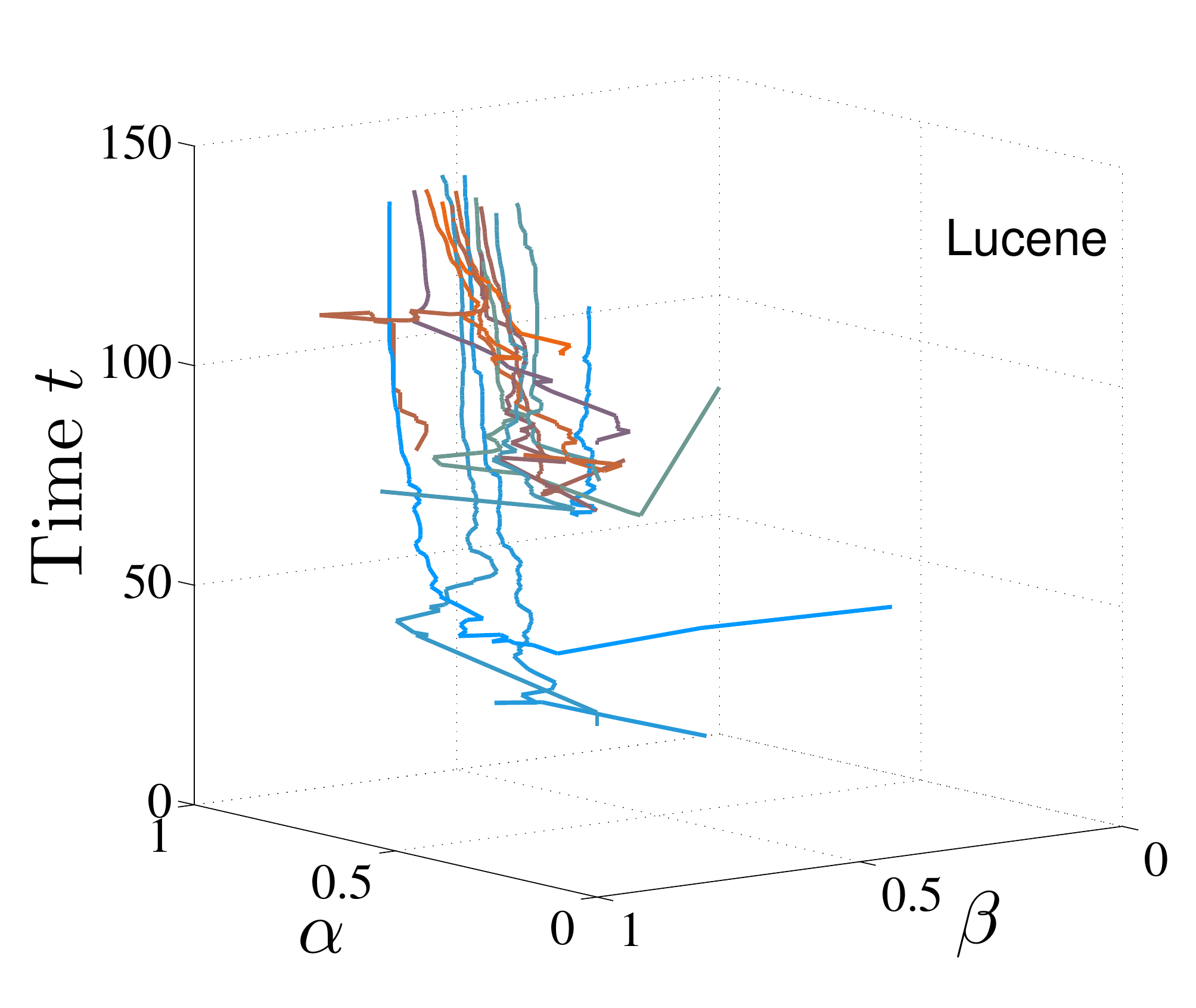}
\caption{Developers' $\alpha$ \& $\beta$ monthly evolving curves, e.g., \emph{Axis2\_java}, \emph{Derby}, and \emph{Lucene}.}
\label{Fig:PDynamics}
\end{figure}

The clustering phenomenon of W-T patterns gives rise to another interesting question: Do developers  choose to join communities with similar W-T patterns as theirs or does the similarity emerge over time as developers participate and evolve with their communities? In the first case, the developers in the same community will present similar W-T patterns from the very beginning, while in the second case, they will get more similar with time. To answer this question, we do the same pattern analysis as above, using only the initial 100 activities in the W-T sequences. Based on the comparison, we find that:
\begin{enumerate}
  \item The developers in the same community showed similar W-T patterns starting with their  inception into the project.
  I.e., for their first 100 activities, the distances of HMM parameters between pairs of developers in the same communities are significantly shorter ($p=3.1\times10^{-13}$) than those from different communities.
  \item In addition, the community cultures of different communities  converge rather than diverge from each other, as time evolves. I.e., both the inner (within-community) and inter (between-community) distances decrease significantly ($p=0$) with time, as shown in Figure~\ref{Fig:DBox}. We also calculate the average inner distance for all communities by considering their respective first $\varrho$ activities with different values of $\varrho$, as shown in Figure~\ref{Fig:PDCurve}, to study the converging process. We find that the inner distances decrease as $\varrho$ increases, for most communities. As examples, the evolutions of the HMM parameters with time for the communities \emph{Axis2\_java}, \emph{Derby}, and \emph{Lucene} are shown in Figure~\ref{Fig:PDynamics}.
  \item The clustering of the HMM parameters within communities grows tighter with time. I.e., the convergence rates of the parameter distances from the first 100 activities to all activities within communities (the average distance decreases from 0.3381 to 0.1832) is significantly larger ($p=1.7\times10^{-7}$) than those between communities (it decreases from 0.4216 to 0.2861).
\end{enumerate}

These findings suggest that developers with similar W-T patterns are indeed more likely to join in the same communities, and continue to harmonize their patterns as they work and talk as a team. In fact, since there are many online communities on similar topics, people can first experience the culture of these communities and then decide to join or not~\cite{cothrel2000measuring,preece2004the,ridings2004virtual}. For OSS, it is clear that most developers do communicate a fair bit on the developer mailing list before actually contributing work~\cite{bird2007open,von2003community}; indeed, this type of ``socialization'' is a necessary pre-requisite to having one's work contributions accepted. Thus, it is to be expected that the developers are more likely to join in the communities with harmonized work and talk patterns, in order to reduce co-ordination efforts.

\begin{table*}[!t]
\centering
\begin{tabular}{|c|c|c|c|c|}
\hline
\multirow{2}*{Property} & \multirow{2}*{Description}
& \multicolumn{3}{c|}{Inter cluster T-test ($p$)}\\
\cline{3-5}
& & \multicolumn{1}{|c}{\#1 vs. \#2} & \multicolumn{1}{|c}{\#2 vs. \#3} & \multicolumn{1}{|c|}{\#1 vs. \#3}\\
\hline
$X_1$ & \multicolumn{1}{|l}{Working rhythm: the number of work activities per day} & \multicolumn{1}{|c}{$1.9\times10^{-2}$} & \multicolumn{1}{|c}{$5.4\times10^{-3}$} & \multicolumn{1}{|c|}{$1.2\times10^{-7}$}\\
\hline
$X_2$ & \multicolumn{1}{|l}{The kilo lines of added codes (KLoC) per day} & \multicolumn{1}{|c}{$1.5\times10^{-2}$} & \multicolumn{1}{|c}{$7.8\times10^{-1}$} & \multicolumn{1}{|c|}{$3.3\times10^{-2}$}\\
\hline
$X_3$ & \multicolumn{1}{|l}{Talking rhythm: the number of talk activities per day} & \multicolumn{1}{|c}{$1.2\times10^{-2}$} & \multicolumn{1}{|c}{$5.8\times10^{-2}$} & \multicolumn{1}{|c|}{$6.6\times10^{-6}$}\\
\hline
$X_4$ & \multicolumn{1}{|l}{The number of new social links per week} & \multicolumn{1}{|c}{$9.0\times10^{-1}$} & \multicolumn{1}{|c}{$3.7\times10^{-2}$} & \multicolumn{1}{|c|}{$2.0\times10^{-3}$}\\
\hline
$X_5$ & \multicolumn{1}{|l}{The observed survival time (year)} & \multicolumn{1}{|c}{$3.0\times10^{-1}$} & \multicolumn{1}{|c}{$3.7\times10^{-2}$} & \multicolumn{1}{|c|}{$2.7\times10^{-1}$}\\
\hline
\end{tabular}
\caption{The student's t-tests for five individual properties between different clusters.}
\label{Tab:BasicMetrics}
\end{table*}

\begin{figure*}
\centering
\centerline{\includegraphics[width=\columnwidth]{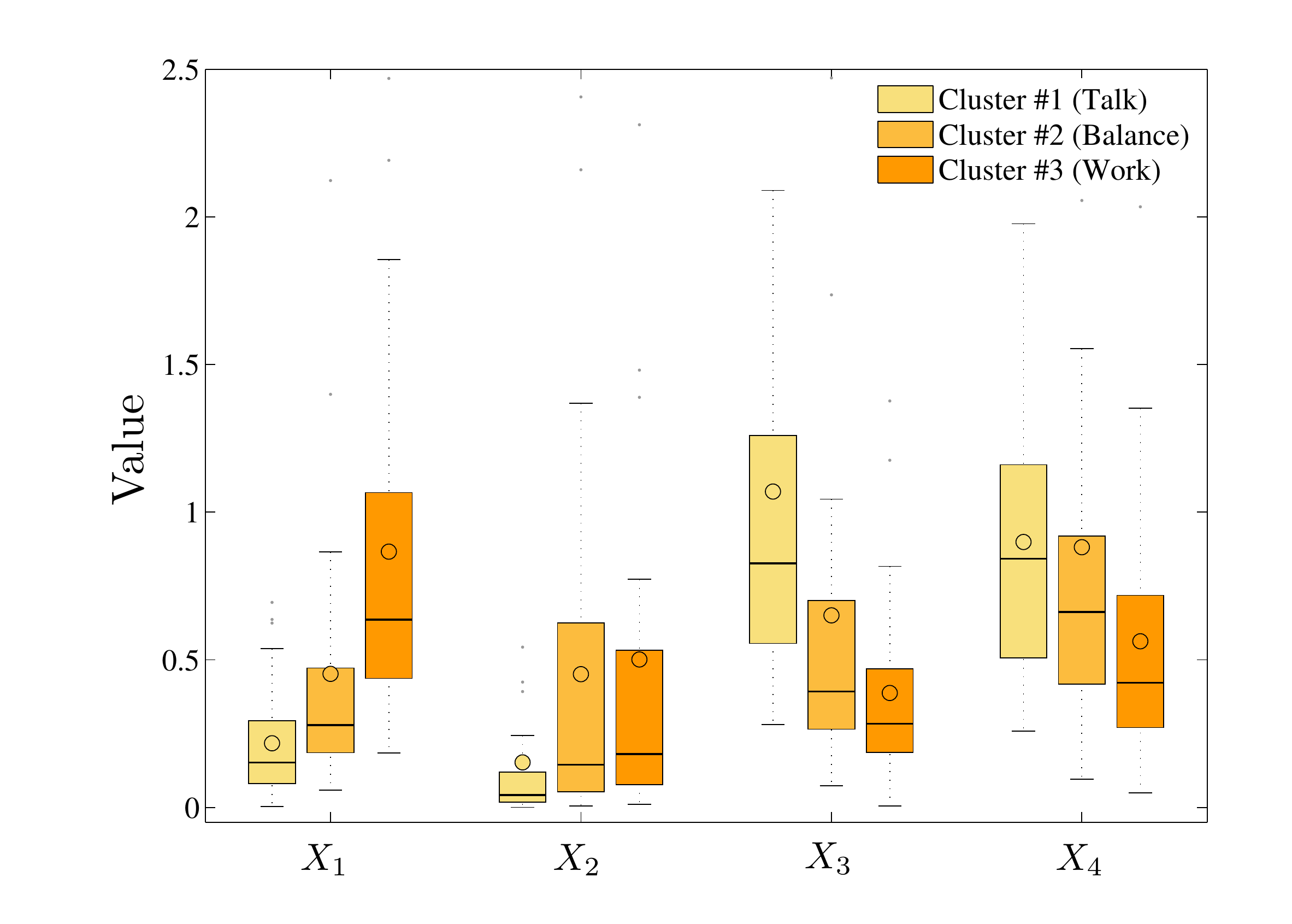}
\includegraphics[width=\columnwidth]{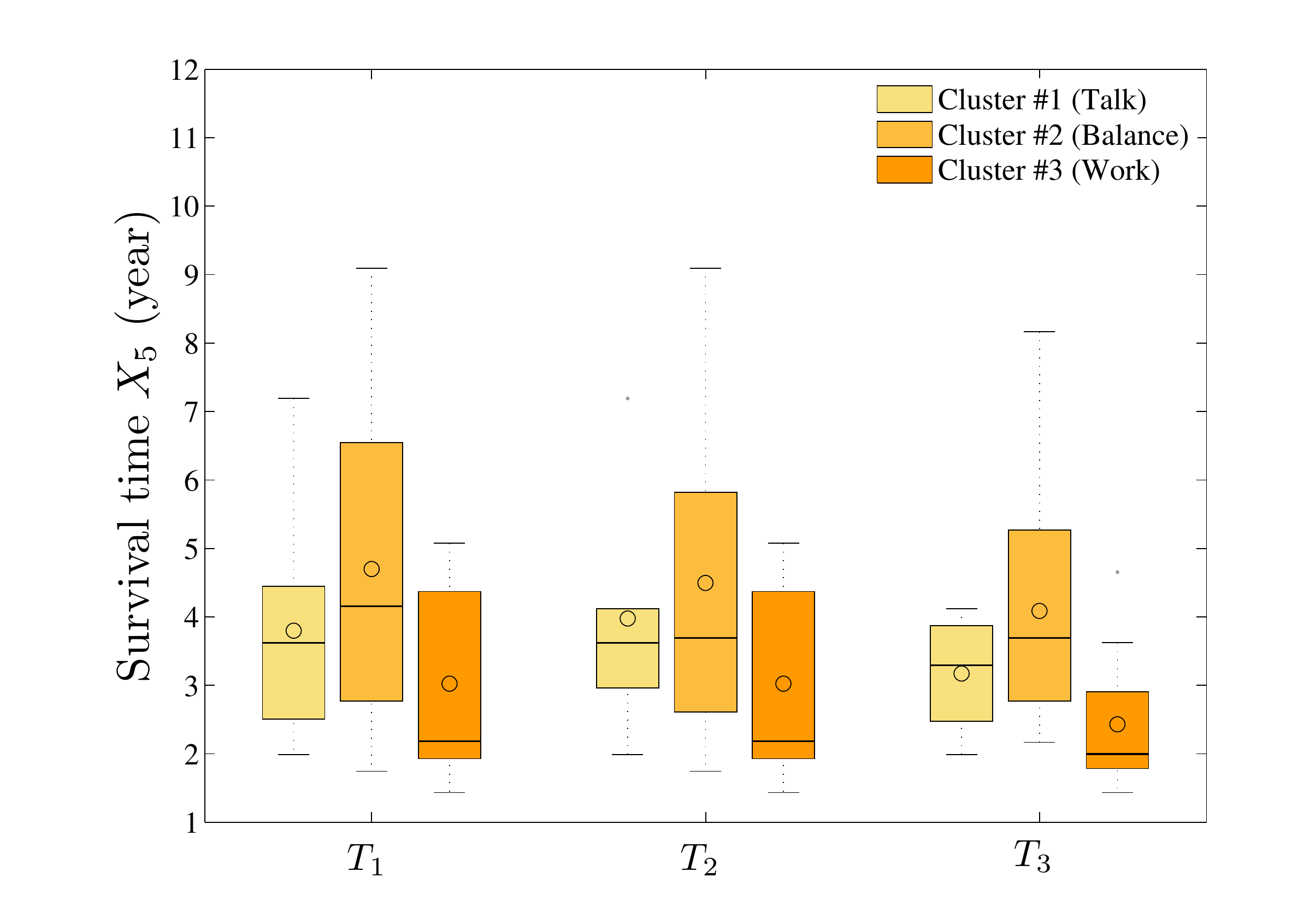}}
\caption{The effects of community culture on individual properties. The box-and-whisker diagrams for (left) the four individual properties $X_1$ to $X_4$, and (right) the observed survival time $X_5$ with different time thresholds $T_1$ (half year), $T_2$ (one year), and $T_3$ (two years), for the developers in the three clusters determined by their HMM parameters. }
\label{Fig:PropertyCluster}
\end{figure*}

In addition, we find that different community cultures will slightly converge rather than diverge from each other over time; this suggests that there may be an over-arching trend of the W-T patterns for all the developers (in all projects). To investigate this further, we compare the two parameters $\alpha$ and $\beta$ separately for all developers, considering \emph{a)} the first 100 activities and \emph{b)} all activities. We find that both of them increase as time evolves, i.e., the HMMs in case \emph{a)} have significantly smaller $\alpha$ ($p=2.7\times{10^{-2}}$) and $\beta$ ($p=1.4\times{10^{-5}}$) than those in \emph{b)}. In fact, the efficiency of overall work and talk activities may be  measured by the sum $\alpha+\beta$; larger values of this sum indicate less switching between activities and thus fewer interruptions. This arguably represents higher efficiency \cite{adamczyk2004if,basoglu2009investigating,latoza2006maintaining}. In other words, the HMM parameters $(\alpha_i,\beta_i)$ shown in Figure~\ref{Fig:PFigure} can be fitted by the linear function:
\begin{equation}
\alpha+\beta=\varepsilon,\label{Eq:LinearFit}
\end{equation}
with a single parameter $\varepsilon$ representing the average efficiency of all the developers. Using the least squares method, we get the average efficiency $\varepsilon$ and the corresponding standard deviation $\sigma$ from the regression line as
\begin{equation}
\varepsilon=\frac{\Sigma_{i=1}^N(\alpha_i+\beta_i)}{N},\quad \sigma=\sqrt{\frac{\Sigma_{i=1}^N(\alpha_i+\beta_i-\varepsilon)^2}{2N}},\label{Eq:LinearFit:PE}
\end{equation}
respectively, for the $N$ developers. We find that the average efficiency steadily increases, while the variance decreases, with time, which means that as time goes on developers tend to have longer bursts of pure work and pure talk, suggesting that their discussions are becoming more effective, and that the ensuing co-operative work proceeds relatively more uninterruptedly.

Looking at the change in the rate of talk activities for all developers, in terms of $\alpha$ and $\beta$, equation~\eqref{Eq:Markov:SteayState}, we find that the rate increases significantly ($p=4.6\times{10^{-3}}$) with time, indicating that most developers become more socialized in the process. This phenomenon is consistent with the fact that more discussions are always needed to further improve a mature product. Meanwhile, contributing to these online communities is social work, i.e., the contributions of developers are highly visible and will be checked by many other users, so it is not surprising that they need to reply to comments more frequently when contributing more.

\subsection{\label{Function}Individual Performance and Community Culture}
Regarding community culture, it is always very interesting to try to identify its benefits on individuals in the respective communities. For example, it is reasonable to hypothesize that developers who work more than they talk will have higher productivity, meaning they will produce more lines of codes (LoC), than those seeking balance between work and talk activities (similarly, those preferring talk over work may have a socialization advantage  compared to those seeking balanced activities). We ask, does increasing productivity always come at the price of decreasing socialization, and vice versa? Note that looking simply for strong work or talk preference, i.e., larger $\alpha$ or $\beta$, respectively, does not necessarily lead to higher productivity or socialization because our sequence analysis does not take the activity time into consideration and is also independent from the length of the sequences.

To answer that we study the correlations with community culture of five measures of individual performance \emph{work rhythm} (\# work activities per day), \emph{thousands of lines of code added per unit time} (KLoC), \emph{talk rhythm} (\# talk activities per day), \emph{newly established social links per week}, and \emph{observed survival time}, resp., $X_1$ to $X_5$, as summarized in Table~\ref{Tab:BasicMetrics}. The first four properties are calculated in the same time period of the person's W-T sequence. The \emph{survival time}, $X_5$, of a developer is defined as the period of time from their first activity to the last one, which may be longer than the period of their W-T sequence, considering that the W-T sequences under study were preprocessed by removing prefixes of pure work or talk activities. The survival time of a developer is only observed when the developer has left the respective community. Here, as a reasonable estimation, we consider that a developer has left the community if they have not been active for a relatively long time, i.e., longer than some threshold $T$.

All developers are divided into three clusters by their HMM parameters, as shown in Figure~\ref{Fig:PFigure}. The developers in Cluster \#1 emphasize ``talk'', those in Cluster \#3 emphasize ``work'', while those in Cluster \#2 seek balance between the two. For each property from $X_1$ to $X_4$, we have a list of their values for developers in each cluster, and the comparisons between the properties of developers in different clusters are visualized by the box-and-whisker diagrams shown in Figure~\ref{Fig:PropertyCluster} (left), with the significance presented in Table~\ref{Tab:BasicMetrics}. We find that the developers in Cluster \#3 have the fastest working rhythms, those in Cluster \#2 follow, while the developers in Cluster \#1 work the slowest. The direction reverses for their talking rhythms. However, the situation is a little different when we compare the abilities of developers of different clusters in producing codes and earning social status. We find that the developers in Cluster \#2 and Cluster \#3 can produce similar KLoC per day, and both groups produce significantly more than the developers in Cluster \#1, while the developers in Cluster \#2 and Cluster \#1 earn similar numbers of social links per week, and both groups earn significantly more than the developers in Cluster \#3. These indicate that extended discussion is always accompanied with the slowing down of work rhythms, but not always with decrease of productivity, and the developers seeking balance between work and talk behave competitively on both productivity and socialization as those who mostly work or mostly talk.

Although it seems that the developers who mostly work have the fastest working rhythms and the highest productivity, on average, it doesn't mean that choice is the healthiest for them or for the overall project, since these developers are more likely to feel boring and then quit the projects. To analyze the survival times of developers (time from joining until leaving) in terms of the HMM parameters $\alpha$ and $\beta$, we use the Hazard model~\cite{elandt1980survival}, with the Hazard ratio defined as
\begin{equation}
\eta=e^{b{x}}.\label{Eq:Hazard:Ratio}
\end{equation}
where $x$ is either $\alpha$ or $\beta$. (See the Appendix for further mathematical descriptions of the Hazard model.) We find that developers with smaller $\alpha$ or larger $\beta$ will have suggestively longer survival times ($p=0.077$ and $b=1.7$ for $\alpha$ and $p=0.042$ and $b=-2.4$ for $\beta$), indicating that, by comparison, talk activities are more important than work activities for developer retention. Indeed, we find that developers with more balance between their work and talk stay active in the projects for suggestively longer periods of time than those who mostly work, as shown in Figure~\ref{Fig:PropertyCluster} (right), i.e., the significance is equal to 0.037, 0.078, and 0.049 when the survival times of the developers with their last activities occurred half year, one year, and two years before are considered, respectively. The significance of comparison for the survival time among the three clusters of developers are presented in Table~\ref{Tab:BasicMetrics} when $T=1$ (year). These findings suggest that developers with balanced W-T patterns are important to sustain OSS communities. Each of the communities we studied has at least one balanced developer, and there is also a natural trend that developers become more balanced, i.e., both $\alpha$ and $\beta$ increase with time.

\begin{table*}[!t]
\centering
\begin{tabular}{|l|c|c|c|c|c|c|c|c|}
\hline
& \multicolumn{4}{c|}{Social weights} & \multicolumn{4}{c|}{Cooperative weights}\\
\cline{2-9}
\multicolumn{1}{|c|}{Communities} & \multicolumn{2}{c|}{Pearson} & \multicolumn{2}{c|}{Spearman} &  \multicolumn{2}{c|}{Pearson} & \multicolumn{2}{c|}{Spearman}\\
\cline{2-9}
& \multicolumn{1}{c|}{$R$} & \multicolumn{1}{c|}{$p$} & \multicolumn{1}{c|}{$R$} & \multicolumn{1}{c|}{$p$} & \multicolumn{1}{c|}{$R$} & \multicolumn{1}{c|}{$p$} & \multicolumn{1}{c|}{$R$} & \multicolumn{1}{c|}{$p$}\\
\hline
Activemq & \multicolumn{1}{c|}{-0.3287} & \multicolumn{1}{c|}{$2.3\times10^{-1}$} & \multicolumn{1}{c|}{-0.3056} & \multicolumn{1}{c|}{$2.7\times10^{-1}$} & \multicolumn{1}{c|}{-0.4427} & \multicolumn{1}{c|}{$9.9\times10^{-2}$} & \multicolumn{1}{c|}{-0.5607} & \multicolumn{1}{c|}{$3.2\times10^{-2}$}\\
\hline
Ant & \multicolumn{1}{c|}{0.0826} & \multicolumn{1}{c|}{$6.3\times10^{-1}$} & \multicolumn{1}{c|}{0.0049} & \multicolumn{1}{c|}{$9.8\times10^{-1}$} & \multicolumn{1}{c|}{-0.3753} & \multicolumn{1}{c|}{$2.4\times10^{-2}$} & \multicolumn{1}{c|}{-0.3704} & \multicolumn{1}{c|}{$2.7\times10^{-2}$}\\
\hline
Axis2\_c & \multicolumn{1}{c|}{-0.4833} & \multicolumn{1}{c|}{$5.8\times10^{-2}$} & \multicolumn{1}{c|}{-0.4667} & \multicolumn{1}{c|}{$1.2\times10^{-2}$} & \multicolumn{1}{c|}{0.2209} & \multicolumn{1}{c|}{$2.6\times10^{-1}$} & \multicolumn{1}{c|}{0.2474} & \multicolumn{1}{c|}{$2.0\times10^{-1}$}\\
\hline
Axis2\_java & \multicolumn{1}{c|}{-0.1611} & \multicolumn{1}{c|}{$1.0\times10^{-1}$} & \multicolumn{1}{c|}{-0.0442} & \multicolumn{1}{c|}{$6.5\times10^{-1}$} & \multicolumn{1}{c|}{-0.0797} & \multicolumn{1}{c|}{$4.2\times10^{-1}$} & \multicolumn{1}{c|}{-0.1714} & \multicolumn{1}{c|}{$8.1\times10^{-2}$}\\
\hline
Camel & \multicolumn{1}{c|}{-0.4896} & \multicolumn{1}{c|}{$6.4\times10^{-2}$} & \multicolumn{1}{c|}{-0.6000} & \multicolumn{1}{c|}{$2.0\times10^{-2}$} & \multicolumn{1}{c|}{-0.2424} & \multicolumn{1}{c|}{$3.8\times10^{-1}$} & \multicolumn{1}{c|}{-0.3679} & \multicolumn{1}{c|}{$1.8\times10^{-1}$}\\
\hline
Cxf & \multicolumn{1}{c|}{-0.0036} & \multicolumn{1}{c|}{$9.9\times10^{-1}$} & \multicolumn{1}{c|}{0.0651} & \multicolumn{1}{c|}{$7.8\times10^{-1}$} & \multicolumn{1}{c|}{0.0711} & \multicolumn{1}{c|}{$7.6\times10^{-1}$} & \multicolumn{1}{c|}{0.1948} & \multicolumn{1}{c|}{$4.0\times10^{-1}$}\\
\hline
Derby & \multicolumn{1}{c|}{-0.2258} & \multicolumn{1}{c|}{$1.3\times10^{-2}$} & \multicolumn{1}{c|}{-0.1940} & \multicolumn{1}{c|}{$3.4\times10^{-2}$} & \multicolumn{1}{c|}{-0.2430} & \multicolumn{1}{c|}{$7.5\times10^{-3}$} & \multicolumn{1}{c|}{-0.3232} & \multicolumn{1}{c|}{$3.4\times10^{-4}$}\\
\hline
Lucene & \multicolumn{1}{c|}{-0.3856} & \multicolumn{1}{c|}{$1.6\times{10^{-4}}$} & \multicolumn{1}{c|}{-0.6046} & \multicolumn{1}{c|}{$2.2\times10^{-10}$} & \multicolumn{1}{c|}{-0.1361} & \multicolumn{1}{c|}{$2.0\times10^{-1}$} & \multicolumn{1}{c|}{-0.2275} & \multicolumn{1}{c|}{$3.0\times10^{-2}$}\\
\hline
Mahout & \multicolumn{1}{c|}{0.6829} & \multicolumn{1}{c|}{$5.0\times{10^{-3}}$} & \multicolumn{1}{c|}{0.6685} & \multicolumn{1}{c|}{$6.4\times10^{-3}$} & \multicolumn{1}{c|}{0.1220} & \multicolumn{1}{c|}{$6.7\times10^{-1}$} & \multicolumn{1}{c|}{-0.3429} & \multicolumn{1}{c|}{$2.1\times10^{-1}$}\\
\hline
Nutch & \multicolumn{1}{c|}{-0.2265} & \multicolumn{1}{c|}{$4.2\times{10^{-1}}$} & \multicolumn{1}{c|}{-0.2832} & \multicolumn{1}{c|}{$3.1\times10^{-1}$} & \multicolumn{1}{c|}{0.3648} & \multicolumn{1}{c|}{$1.8\times10^{-1}$} & \multicolumn{1}{c|}{0.4071} & \multicolumn{1}{c|}{$1.3\times10^{-1}$}\\
\hline
Ode & \multicolumn{1}{c|}{-0.4722} & \multicolumn{1}{c|}{$7.6\times{10^{-2}}$} & \multicolumn{1}{c|}{-0.4866} & \multicolumn{1}{c|}{$6.6\times10^{-2}$} & \multicolumn{1}{c|}{-0.1250} & \multicolumn{1}{c|}{$6.6\times10^{-1}$} & \multicolumn{1}{c|}{-0.1429} & \multicolumn{1}{c|}{$6.1\times10^{-1}$}\\
\hline
Openejb & \multicolumn{1}{c|}{-0.0017} & \multicolumn{1}{c|}{1.0} & \multicolumn{1}{c|}{0.0667} & \multicolumn{1}{c|}{$8.6\times10^{-1}$} & \multicolumn{1}{c|}{-0.6465} & \multicolumn{1}{c|}{$4.3\times10^{-2}$} & \multicolumn{1}{c|}{-0.3818} & \multicolumn{1}{c|}{$2.8\times10^{-1}$}\\
\hline
Solr & \multicolumn{1}{c|}{-0.3188} & \multicolumn{1}{c|}{$9.8\times{10^{-2}}$} & \multicolumn{1}{c|}{-0.5083} & \multicolumn{1}{c|}{$5.8\times10^{-3}$} & \multicolumn{1}{c|}{-0.5426} & \multicolumn{1}{c|}{$2.9\times10^{-3}$} & \multicolumn{1}{c|}{-0.5457} & \multicolumn{1}{c|}{$3.1\times10^{-3}$}\\
\hline
Wicket & \multicolumn{1}{c|}{-0.2512} & \multicolumn{1}{c|}{$2.0\times{10^{-1}}$} & \multicolumn{1}{c|}{-0.1363} & \multicolumn{1}{c|}{$4.9\times10^{-1}$} & \multicolumn{1}{c|}{-0.2326} & \multicolumn{1}{c|}{$2.3\times10^{-1}$} & \multicolumn{1}{c|}{-0.1795} & \multicolumn{1}{c|}{$3.6\times10^{-1}$}\\
\hline
All & \multicolumn{1}{c|}{-0.2396} & \multicolumn{1}{c|}{$1.6\times{10^{-8}}$} & \multicolumn{1}{c|}{-0.2517} & \multicolumn{1}{c|}{$2.8\times10^{-9}$} & \multicolumn{1}{c|}{-0.1420} & \multicolumn{1}{c|}{$9.1\times10^{-4}$} & \multicolumn{1}{c|}{-0.2037} & \multicolumn{1}{c|}{$1.8\times10^{-6}$}\\
\hline
\end{tabular}
\caption{The Pearson \& Spearman correlation tests for the distances of HMM parameters and social \& cooperative weights between pairwise developers for different communities.}
\label{Tab:CorrTest}
\end{table*}

\section{\label{NetworkEffect}Synchronizing Role of Social and Technical Links}
It is well understood that individuals can influence each other's behaviors through social links~\cite{borgatti2009network,centola2010the,fowler2010cooperative}. Here, we study the extent to which developers with similar W-T patterns tend to be linked more in the email network or the technical cooperation network. In social networks, social weight between two developers intuitively means the number of emails between them. In cooperation networks, a pair of developers are linked with an edge indicating the number of files on which they have both worked.
In particular, denoting by $\psi_i$ the list of files that developer $d_i$ commits to, the cooperative weight between a pair of developers $d_i$ and $d_j$, in terms of the files to which they have committed, is defined as
\begin{equation}
\omega_{ij}=\frac{\psi_i\cap\psi_j}{\psi_i\cup\psi_j}. \label{Eq:Similarity}
\end{equation}

On the social end, for pairs of developers, we ask: are the distances between their HMM parameters correlated with the number of emails they have exchanged? The results of using both Pearson and Spearman correlations are given in Table~\ref{Tab:CorrTest}, under the Social weights columns. We find negative correlation in ten out of fourteen communities with both methods, with the significance $p<0.05$ in eight of them, while we find positive correlation with the significance $p<0.1$ in only one community called \emph{Mahout}. The negative correlation means that the smaller the HMM parameter distance between two developers, the larger the number of emails they have exchanged.

On the technical end, we study the correlation between the distances of HMM parameters and strength of file cooperation links between developers. Using the same correlation measures as before, we get the results in Table~\ref{Tab:CorrTest}, under the Cooperative weights columns. In this case negative correlation is found in eleven out of fourteen communities with both methods, with the significance $p<0.1$ in four of them, including \emph{Activemq}, \emph{Ant}, \emph{Derby}, and \emph{Solr}, while no community has positive correlation with significance $p<0.1$. The negative correlation means that the smaller the HMM parameter distance between two developers, the larger the cooperation between them.

When considering all communities together, we obtain a significantly negative correlation for both methods in both cases (the last row of Table~\ref{Tab:CorrTest}). Thus, developers with more emails between them or committing to more of the same files are more likely to have similar W-T patterns. The results also indicate that community culture may be either social or task oriented (technical); the distances between HMM parameters are more likely to be correlated with social weights in some communities, and with cooperative weights in others.

\section{\label{Threats}Threats to Validity}
Although most results shown in this paper are significant, there are still a number of threats to this study. 

All the OSS projects under study are collected from Apache, which may limit the generalization of the results. The methods therefore need to be tested on other OSS ecosystems, such as GitHub~\cite{dabbish2012social} and GNOME~\cite{sarma2009tesseract}, or other online volunteer communities, such as Wikipedia~\cite{kittur2007he,viegas2007talk} in the future. We have collected some GitHub data, and also find the converging W-T patterns there. However, the correlation between W-T patterns and productivity of developers need to be further validated, since currently we haven't yet collected the data about code length added or deleted in a particular commit. We will show the extended results in our future work.

We acknowledge that, like many other empirical researches, our work is based on a sample of work and talk activities of developers, but not all of them. We just consider the commits to code or documents as work activities, while in reality developers may have other kinds of work activities which are relatively difficult to be captured from OSS repositories. Besides, there might be other kinds of talk activities too, e.g., the discussion on issue tracking systems, that are not included in this study. In fact, we indeed collect issue tracking data from Jira and Bugzilla~\cite{gharehyazie2013social}, and do experiments by including them as talk activities (both opening issue as initializing the discussion and comments). However, we don't find any result that changes dramatically in this case, indicating the revealed phenomena are quite robust.          

We use only LoC to measure the productivity of a developer, while in fact there are alternative metrics, such as the number of issues fixed~\cite{zhou2012make}. Moreover, there are also metrics about work efficiency, such as the development time of tasks~\cite{herbsleb2006collaboration}, and about the quality of code, such as the number of bugs~\cite{aberdour2007achieving}. We will study the effects of work-talk patterns on these metrics in the future, rather than put them together in this single paper. It's reasonable to use LoC here, since it has been used extensively to measure the productivity of developers~\cite{mockus2002two}. 

\section{\label{Conclusion}Conclusions}
In this paper, we demonstrate that work-talk patterns of software developers in a number of OSS communities can be effectively studied using sequence analysis methods on sequences arising from simple two-state behavior models.
Our methods enabled us to learn about a series of interesting task-oriented community based phenomena: that developers in a community present similar W-T patterns, and this clustering of W-T patterns is enhanced with time, reflecting different work cultures in these communities, with emphasis on different proportions of continuous work to continuous talk activities; that social and technical interactions may play a role in synchronizing W-T patterns, since developers with stronger social or technical links in a community have more similar W-T patterns; and that although successful task-communities may have relatively different cultures, developers with balanced work-talk patterns seems to play critical roles in sustaining them, and, at least in the ones we studied, each has at least one such developer.

These findings suggest that online individuals may synchronize their behaviors with others to better fit in the task communities and to improve coordinating efficiencies. The methods proposed in this paper can be further expanded and applied to analyze the switching pattern of more varied kinds of activities in more diverse online communities.

\section{Acknowledgments}
The authors would like to thank all the members in our research group in the Department of Computer Science, University of California in Davis, for the valuable discussion about the ideas and technical details presented in this paper. All authors gratefully acknowledge support from the Air Force Office of Scientific Research, award FA955-11-1-0246. QX acknowledges support from the National Natural Science Foundation of China (Grant Nos. 61004097, 61273212) and the China Scholarship Council (CSC).

\balance

\small{
\bibliographystyle{abbrv}
\bibliography{sample}
}

\section{Appendix}

\subsection{Hidden Markov Model}
An HMM is a simple modeling mechanism to explain transitions among several different states. We use an HMM with two states,``work'' and ``talk'', and transitions between them corresponding to either continuing to perform the same activity, W followed by a W or T followed by a T, or switching activities, W followed by a T, and vice versa.
The HMM diagram is shown in Figure~\ref{Fig:MarkovModel}. If we denote by $P_\textrm{W}(k)$ and $P_\textrm{T}(k)$ the probabilities that work, resp. talk, happen at time step $k$, then for the next time point we have
\begin{eqnarray}
P_\textrm{W}(k+1)&=&\alpha{P_\textrm{W}(k)}+(1-\beta)P_\textrm{T}(k),\label{Eq:Markov:Work}\\
P_\textrm{T}(k+1)&=&(1-\alpha)P_\textrm{W}(k)+{\beta}P_\textrm{T}(k).\label{Eq:Markov:Talk}
\end{eqnarray}
where $\alpha$ and $\beta$ are the transition probabilities.
We note here that while $\alpha$ and $\beta$ could evolve with time, they don't change much between successive activities, therefore we can consider them as constants in the sequences with certain lengths.

Equations \eqref{Eq:Markov:Work} and \eqref{Eq:Markov:Talk} can be approximated for continuous time, $\tau$, and then transformed to the following more compact matrix form:
\begin{equation}
\dot{P}(\tau)=\left[
  \begin{array}{cc}
    \alpha-1 & 1-\beta\\
    1-\alpha & \beta-1\\
  \end{array}
\right]P(\tau),\label{Eq:Markov:Matrix}
\end{equation}
with $P(\tau)=[P_\textrm{W}(\tau),P_\textrm{T}(\tau)]^\mathrm{T}$. By solving equation \eqref{Eq:Markov:Matrix}, we have
\begin{equation}
P(\tau)=C_1\left[
  \begin{array}{c}
    1-\beta\\
    1-\alpha\\
  \end{array}
\right]+C_2\left[
  \begin{array}{c}
    1\\
    -1\\
  \end{array}
\right]e^{(\alpha+\beta-2)\tau}.\label{Eq:Markov:Solution}
\end{equation}

The fractions of work and talk activities, $P_\textrm{W}$ and $P_\textrm{T}$, in a sequence with length $L$ can be estimated by
\begin{eqnarray}
\left[
  \begin{array}{c}
    P_\textrm{W}\\
    P_\textrm{T}\\
  \end{array}
\right]&=&\frac{1}{L}\int_{0}^{L}P(\tau).\label{Eq:Markov:Integral}
\end{eqnarray}

By substituting equation~\eqref{Eq:Markov:Solution} into ~\eqref{Eq:Markov:Integral}, we have
\begin{eqnarray}
\left[
  \begin{array}{c}
    P_\textrm{W}\\
    P_\textrm{T}\\
  \end{array}
\right]&=&\frac{C_1}{(2-\alpha-\beta)L}\left[
  \begin{array}{c}
    1\\
    -1\\
  \end{array}
\right]\left[1-e^{(\alpha+\beta-2)L}\right]\nonumber \\ 
&+&C_2\left[
  \begin{array}{c}
    1-\beta\\
    1-\alpha\\
  \end{array}
\right] .\label{Eq:Markov:IntegralSolution}
\end{eqnarray}

In the right side of equation~\eqref{Eq:Markov:IntegralSolution}, the first term is negligible when the sequence is long enough, considering $\alpha+\beta<2$. Since it is always satisfied $P_\textrm{W}+P_\textrm{T}=1$, we have
\begin{equation}
P_\textrm{W}=\frac{1-\beta}{2-\alpha-\beta}, \quad P_\textrm{T}=\frac{1-\alpha}{2-\alpha-\beta},\label{Eq:Markov:SteayState}
\end{equation}
which are fully determined by the two parameters in the model. Then, the probabilities for the four different two-patterns in the sequence, in terms of $\alpha$ and $\beta$, are given by:
\begin{eqnarray}
P_{\textrm{WW}}&=&\alpha{P_\textrm{W}}=\frac{\alpha(1-\beta)}{2-\alpha-\beta},\label{Eq:Pattern:WW}\\
P_{\textrm{WT}}&=&(1-\alpha)P_\textrm{W}=\frac{(1-\alpha)(1-\beta)}{2-\alpha-\beta},\label{Eq:Pattern:WT}\\
P_{\textrm{TW}}&=&(1-\beta)P_\textrm{T}=\frac{(1-\alpha)(1-\beta)}{2-\alpha-\beta},\label{Eq:Pattern:TW}\\
P_{\textrm{TT}}&=&\beta{P_\textrm{T}}=\frac{(1-\alpha)\beta}{2-\alpha-\beta},\label{Eq:Pattern:TT}
\end{eqnarray}
Intuitively, larger $\alpha$ and $\beta$ means higher proportions of WW and TT patterns, respectively, in the sequence. Furthermore, the probabilities for longer patterns can be calculated similarly, once the model parameters $\alpha$ and $\beta$ are estimated from equations~\eqref{Eq:Pattern:WW} to \eqref{Eq:Pattern:TT}. It is important to note that we always have $\alpha+\beta=1$ in the randomized W-T sequences generated by the null model. In this case, $\alpha$ and $\beta$ are equal to the fractions of work and talk activities, respectively.

\subsection{Hazard Model}
Survival analysis enables modeling of outcomes in the presence of censored data.
In our case the censoring is due to the uncertainty that long time periods without activities may or may not indicate a developer has left the community.
Generally, survival analysis involves calculating the Hazard rate, defined as the limit of the number of events per $\delta{t}$ time divided by the number at risk, as $\delta{t}\rightarrow{0}$.
Here, suppose a developer does not leave the community until time $\Gamma$, the Hazard rate is given by
\begin{equation}
h(t)=\lim_{\delta{t}\rightarrow0}\frac{P(t\leq{\Gamma}<t+\delta{t}|t\leq{\Gamma})}{\delta{t}}.\label{Eq:Hazard:Ratio}
\end{equation}
Our primary interest is the survival function defined as $S(t)=P(t<\Gamma)$, which can be calculated from equation \eqref{Eq:Hazard:Ratio} by
\begin{equation}
S(t)=e^{-\int_0^th(\tau)d\tau}.\label{Eq:Hazard:General}
\end{equation}
Suppose there are several factors, denoted by $x_i,i=1,2,\ldots,\gamma$, that can influence the survival time, then we adopt the Cox model~\cite{cox1972regression} to define the Hazard rate $h(t)$ by
\begin{equation}
h(t)=h_0(t)e^{\sum_{i=1}^\gamma{b_i{x_i}}},\label{Eq:CoxModel}
\end{equation}
with $h_0(t)$ describing how the hazard changes over time at baseline levels of covariates. Here we focus on the hazard ratio $h(t)/h_0(t)$ to see whether increasing some covariates will significantly increase or decrease the survival time, e.g., $b_i>0$ means that the individuals of larger $x_i$ will have statistically shorter survival times.
\end{document}